\documentclass[sigconf]{acmart}
\AtBeginDocument{%
  }

\setcopyright{none}
\definecolor{darkgreen}{rgb}{0.0, 0.5, 0.0}

\usepackage{dirtytalk}
\usepackage{subcaption}
\usepackage{multirow}
\usepackage{tabularx}
\usepackage{enumitem}

\usepackage{amsmath}
\DeclareMathOperator*{\argmin}{arg\,min}

\usepackage{xcolor}
\definecolor{powerpointgreen}{RGB}{197, 224, 180}
\definecolor{powerpointblue}{RGB}{132,151,176}
\definecolor{powerpointorange}{RGB}{244,177,131}

\usepackage{tikz}
\newcommand*\circledblue[1]{\tikz[baseline=(char.base)]{
            \node[shape=circle,draw=powerpointblue!150,fill=powerpointblue,thick,inner sep=1pt] (char) {\textsf#1};}}
\newcommand*\circledorange[1]{\tikz[baseline=(char.base)]{
            \node[shape=circle,draw=powerpointorange!150,fill=powerpointorange,thick,inner sep=1pt] (char) {\textsf#1};}}
\newcommand*\circledgreen[1]{\tikz[baseline=(char.base)]{
            \node[shape=circle,draw=powerpointgreen!150,fill=powerpointgreen,thick,inner sep=1pt] (char) {\textsf#1};}}
            
\begin{document}

\title{Analyzing User Characteristics of Hate Speech Spreaders on Social Media}
% Causal Understanding of Why Users Share Hate Speech on Social Media
% Debiasing hate speech Resharing Behavior
% A debiasing framework for hate speech resharing behavior
% A debiasing framework for analyzing hate speech resharing
\author{Dominique Geissler}
\affiliation{%
  \institution{LMU Munich \& Munich Center for Machine Learning}
  \city{Munich}
  \country{Germany}
}
\email{d.geissler@lmu.de}

\author{Abdurahman Maarouf}
\affiliation{%
  \institution{LMU Munich \& Munich Center for Machine Learning}
  \city{Munich}
  \country{Germany}
}
\email{a.maarouf@lmu.de}

\author{Stefan Feuerriegel}
\affiliation{%
  \institution{LMU Munich \& Munich Center for Machine Learning}
  \city{Munich}
  \country{Germany}
}
\email{feuerriegel@lmu.de}


\begin{abstract}
{Hate speech on social media threatens the mental and physical well-being of individuals and contributes to real-world violence. Resharing is an important driver behind the spread of hate speech on social media. Yet, little is known about \emph{who} reshares hate speech and what their characteristics are. In this paper, we analyze the role of user characteristics in hate speech resharing across different types of hate speech (e.g., political hate). For this, we first cluster hate speech posts using large language models into different types of hate speech. Then we model the effects of user attributes on users' probability to reshare hate speech using an explainable machine learning model. To do so, we apply debiasing to control for selection bias in our observational social media data and further control for the latent vulnerability of users to hate speech. We find that, all else equal, users with fewer followers, fewer friends, fewer posts, and older accounts share more hate speech. This shows that users with little social influence tend to share more hate speech. Further, we find substantial heterogeneity across different types of hate speech. For example, racist and misogynistic hate is spread mostly by users with little social influence. In contrast, political anti-Trump and anti-right-wing hate is reshared by users with larger social influence. Overall, understanding the factors that drive users to share hate speech is crucial for detecting individuals at risk of engaging in harmful behavior and for designing effective mitigation strategies.} \textbf{Disclaimer}: This work contains terms that are offensive and hateful. 
to the nature of the work.
\end{abstract}

\begin{CCSXML}
<ccs2012>
   <concept>
       <concept_id>10003456.10010927</concept_id>
       <concept_desc>Social and professional topics~User characteristics</concept_desc>
       <concept_significance>300</concept_significance>
       </concept>
   <concept>
       <concept_id>10003120.10003130.10003131.10003234</concept_id>
       <concept_desc>Human-centered computing~Social content sharing</concept_desc>
       <concept_significance>500</concept_significance>
       </concept>
   <concept>
       <concept_id>10003120.10003130.10003131.10011761</concept_id>
       <concept_desc>Human-centered computing~Social media</concept_desc>
       <concept_significance>500</concept_significance>
       </concept>
   <concept>
       <concept_id>10010405.10010455.10010461</concept_id>
       <concept_desc>Applied computing~Sociology</concept_desc>
       <concept_significance>300</concept_significance>
       </concept>
 </ccs2012>
\end{CCSXML}

\ccsdesc[300]{Social and professional topics~User characteristics}
\ccsdesc[500]{Human-centered computing~Social content sharing}
\ccsdesc[500]{Human-centered computing~Social media}
\ccsdesc[300]{Applied computing~Sociology}

\keywords{hate speech, social media, propensity score, recommender systems, user characteristics, resharing}


\maketitle

\section{Introduction}

% hate speech as problem
Online hate speech poses a significant problem to society. Around one in four U.S. adults has experienced online hate \cite{Vogels.2021}, and, in particular, young adults suffer from the psychological consequences \cite{Saha.2019}. Hate speech not only threatens users' mental and physical well-being but also has the potential to fuel radicalization and incite real-world violence \cite{Muller.2021}, as was the case for the suspect of the 2018 Pittsburgh synagogue shooting \cite{Roose.2018}.

% motivation
On social media, reshares are vital in disseminating content beyond immediate circles \cite{Subbian.2017}. One risk of reshares is that harmful content can go viral, such as propaganda \cite{Geissler.2023}, fake news \cite{Cheng.2021, Shao.2018, Prollochs.2021, Prollochs.2021b}, and content from untrustworthy sources \cite{Guess.2023}. Prior research found that reshares also make hate speech go disproportionally more viral than normal content \citep{Maarouf.2024, Mathew.2018}, which is exploited by hateful users to reach more people \cite{Goel.2023}.

% RW
However, prior work has mostly looked into \emph{who posts} hate speech \cite{Hua.2020, Ribeiro.2018} and \emph{who is targeted} by it \cite{ElSherief.2018}. Hateful users are characterized by high activity levels, complex word usage, and high connectivity among each other \cite{Hua.2020, Ribeiro.2018}. Similarly, targets of hate speech show high levels of activity, but they also tend to be older accounts and have more followers \cite{ElSherief.2018}. However, no work has been conducted to profile the users \emph{who reshare} hate speech, although reshares help hate speech reach larger audiences and are an important driver to the virality of hate speech \cite{Maarouf.2024}. Hence, in this work, we analyze the effects of attributes (such as, e.g., verified status, account age, or number of posts, followers, and friends) of users who reshare hate speech while taking into account the different types of hate speech.

% empirical challenges
Estimating the effects of user attributes on users' probability to reshare hate speech is difficult. One challenge is the inherent \emph{selection bias} in observational social media click data, which comes from the users' self-selection of which content to interact with as well as the recommendations from online platforms \cite{Schnabel.2016}. For example, users only interact with content that they choose and online platforms only recommend content that they assume a user is likely to engage with. We later employ inverse propensity scoring (IPS) \cite{Rosenbaum.1983} to reweight the observed interactions between users and hate speech posts for debiasing. Another reason is that we have to control for the \emph{varying vulnerability of users to hate speech} as that can cause bias in the estimates of the effects being studied. Previous research argued that past latent vulnerability (PLV) of a user to specific content influences both the user attributes and the probability to reshare \cite{Benevenuto.2009, Cheng.2021}. Users may have varying PLV as some users simply share more hate speech than others. Also, the risk of exposure to hate speech is different for each user, e.g. how many times a user has seen hate speech in their feed. In our setting, we control for PLV to hate speech.

{In this work, we analyze the role of user characteristics in hate speech resharing across different types of hate speech. Our framework follows three steps: \circledgreen{1}~We first cluster hate speech posts into their type using BERTopic \cite{Grootendorst.2022.bertopic} and employ automatic cluster labeling with LLAMA-3 \cite{Llama3.2024}. This way, we obtain a topic label for each hate speech post. \circledblue{2}~We then model the past latent vulnerability of each user to hate speech. For this, we debias the observational data by adapting inverse propensity scoring (IPS) from causal inference \cite{Rosenbaum.1983}. This reweights the likelihood of users to be exposed to hate speech \cite{Liang.2016, Schnabel.2016} and we obtain \emph{debiased} propensity scores. We employ the debiased propensity scores to model the debiased PLV of users to hate speech by learning latent representations in the form of debiased embeddings. \circledorange{3}~We then model the effects of user attributes on the probability to reshare hate speech while controlling for the PLV. Here, we compute the likelihood to reshare each type of hate speech per user by looking at past reshares. Then, we model the (non-linear) effects of the user attributes (verified status, account age, number of posts, followers, and friends) on the user's probability to reshare hate speech in the future using an explainable boosting machine (EBM) \cite{EBM}.\footnote{Codes and data for reproducibility are available via an anonymous repository for review and will be available in a public repository upon publication: \url{https://anonymous.4open.science/r/analyzing_user_characteristics_hatespeech_spreaders-2042/README.md}}}

\section{Related Work}

\subsection{Hate speech on social media}
\label{sec:hatespeech_socialmedia}
Hate speech is defined as \say{abusive speech targeting specific group characteristics, such as ethnic origin, religion, gender, or sexual orientation} \cite{Warner.2012}. As such, hate speech expresses animosity against specific groups and might even call for real-world violence. 
% content level analysis
In general, hateful posts tend to go more viral than normal content since their cascades are larger, live longer, and are of larger structural virality \cite{Maarouf.2024, Mathew.2018}. 

% Posts with higher frequencies of moralized language receive more hateful replies \cite{Solovev.2023}.

% Hate speech detection 
One stream of literature conducted content-level analyses. Here a frequent theme is to develop machine learning models that can detect and subsequentially remove hate speech. Such systems aim to detect hate speech, for example by using large language models \cite{Barbieri.2020, Djuric.2015, Mozafari.2019} or other natural language processing methods \cite{Badjatiya.2017,  Davidson.2017}. 

% user level analysis
Another stream of literature conducted user-level analyses. Targets of hate speech tend to be more popular or high-profile users \cite{ElSherief.2018}, while users that post hate speech are characterized by higher social media activity and tend to act in more connected networks \cite{Hua.2020, Ribeiro.2018}. When looking at hate speech in online political discourse, people of color of the Democratic Party, white Republicans, and women in general tend to be the main targets \cite{Solovev.2022}. 

{However, none of the above works analyze \emph{resharing} of hate speech. Yet, reshares are what makes content go viral and play a vital role in the dissemination of hate speech online. Moreover, no work has been conducted that analyzes \emph{who} reshares hate speech and specifically which \emph{type} of hate speech. This is in contrast to our work, which aims to answer which user attributes make users reshare hate speech and specifically the different types of hate speech.}

\subsection{Inverse propensity scoring}
Click data from observational social media data are subject to selection bias. Specifically, previous research identifies two sources of selection bias: self-selection by the users and recommendations from online platforms \cite{Schnabel.2016}. On the one hand, users self-select the content they interact with on social media. On the other hand, online platforms tend to recommend only content to users, that a user is likely interested in. This means that click data are biased because the observed interactions were  influenced by self-selection and recommendation bias. Hence, we must carefully account for selection bias in our observational click data.

To alleviate selection bias in observational click data, previous works draw upon a popular technique from causal inference, specifically inverse propensity scoring (IPS) \cite{Rosenbaum.1983, Imbens.2015, Rubin.2001}. Each hate speech post is reweighted with its inverse propensity score, i.e. the probability that a user is exposed to the post. This way, we transform the observational click data as though it came from a randomized trial where users are randomly shown items \cite{Liang.2016, Schnabel.2016}. By doing so, the selection bias is alleviated and debiased propensity scores are obtained. Some works have applied IPS to learn and evaluate debiased recommendation models \cite{Liang.2016, Schnabel.2016}. Others have applied IPS to debias observational data used to learn users' fake news resharing behavior on social media \cite{Cheng.2021}. In particular, the estimated propensity scores of fake news were debiased in three ways: (a)~by using the popularity of shared fake news items; (b)~by using the popularity of resharing users; or (c)~by employing a neural network \cite{Cheng.2021}.

% Applying IPS creates a pseudo-population that mimics a randomized trial by reweighting the samples in different treatment groups \cite{Rosenbaum.1983, Rubin.2001}. 

In the context of our work, the propensity scores describe the probability that a user is exposed to hate speech. Here, we also deal with selection bias of our observational click data. Hence, we later also adopt the debiased propensity score estimations and integrate them into our framework for modeling hate speech resharing. This way, we obtain debiased propensity scores for the exposure of users to hate speech, which we can use to model debiased, latent representations of the vulnerability to hate speech. 

\subsection{Modeling past latent vulnerability}
In general, the vulnerability of users to specific online content cannot be directly observed from data but instead can be represented by latent embeddings. Previous work modeled PLV through recommendation algorithms \cite{Bonner.2018}. Recommendation algorithms take into account previous interactions of users with online content in order to suggest new content the user is likely to interact with. The recommendation algorithms are then evaluated by the coherence of recommendation candidates with past user behavior. This way, they inherently mimic user behavior \cite{Bonner.2018}. For example, prior work leveraged recommendation algorithms to model the PLV of users to fake news, which was used as a confounder to find the attributes that cause users to reshare fake news online \cite{Cheng.2021}. However, to the best of our knowledge, no work modeled the PLV of users to hate speech on social media.

We later transfer the idea of recommendation algorithms to model the PLV of users to hate speech. In particular, we employ recommendation algorithms to learn the debiased PLV of users to hate speech from IPS-reweighted data. By doing so, we represent the PLV as a \emph{debiased} embedding.

\section{Methodology}

\subsection{Overview}

Our objective is to model the effects of user attributes $A$ on the users' probability to reshare different types of hate speech $Y$ while accounting for the user's (latent) vulnerability to hate speech $X$. For this, we use BERTopic to cluster posts into different types of hate speech and employ a state-of-the-art, explainable debiasing framework to model the effects. 

\textbf{Input:} We input the interactions of users with hate speech posts as a bipartite graph into our framework. Let $\mathcal{U} = \{u_1, \ldots, u_i\}$ denote users and $\mathcal{H} = \{h_1, \ldots, h_j\}$ denote the hate speech posts. The interaction between user $u \in \mathcal{U}$ and hate speech post $h \in \mathcal{H}$ is then a binary variable denoted as $S_{uh} = \{0,1\}$, where $S_{uh} = 1$ if user $u$ shared hate speech post $h$, otherwise $S_{uh} = 0$.

\protect\circledgreen{1}~{\textbf{Clustering hate speech:} In the first step, we employ BERTopic \cite{Grootendorst.2022.bertopic} combined with HDBSCAN to cluster hate speech posts $\mathcal{H}$. Subsequently, we label the clusters using automatic labeling with LLAMA-3 \cite{Llama3.2024}. Hence, we generate clusters of types of hate speech $\mathcal{C} = \{c_1, \ldots, c_n\}$. }

\protect\circledblue{2}~{\textbf{Modeling past latent vulnerability:} In the next step, we model the past latent vulnerability $X$ of users to hate speech. For this, we first apply IPS reweighting to debias our observational data, which reweights the propensity $\pi$ of hate speech posts to mimic a randomized trial, and, thereby, we address the issue of selection bias from above. Then, we model the past latent vulnerability to hate speech $X_u$ for each user $u$.}

\protect\circledorange{3}~{\textbf{Modeling hate speech resharing:} In the final step, we model the effects of user attributes $A$ on users' probability to reshare hate speech $Y$ while controlling for PLV ${X}$. We write the individual user's probability to reshare hate speech in general as outcome variable $Y_u \in [0,1]$ and the probability to reshare hate speech type $c$ as $Y_{uc} \in [0,1]$. Then, we use an explainable boosting machine (EBM) \cite{EBM} to predict the probability to reshare hate speech. For each user $u$ with $m$ user attributes denoted as vector ${a}_u = (a_1, \ldots, a_m)$, ${a}_u \in A$ and PLV embedding ${X}_u$ we model the probability to reshare hate speech $Y_u$ ($Y_{uc}$ for different types of hate speech $c$).}

\subsection{Clustering hate speech (Step~1)}
{Given hate speech posts $\mathcal{H}$, we employ BERTopic \cite{Grootendorst.2022.bertopic}, which leverages BERT embeddings and topic modeling, combined with HDBSCAN for clustering. First, we remove stop words, URLs, and emojis. Then, we generate embeddings for each hate speech post using the pre-trained BERT model \emph{all-MiniLM-L6-v2} to capture the semantic content. Next, we reduce the embedding dimensionality using UMAP to enhance performance and interpretability. Lastly, we apply HDBSCAN to the reduced embeddings to identify clusters~$\mathcal{C}$. HDBSCAN is highly flexible as it does not require pre-specified numbers of clusters and it is suitable for noisy data, which makes it an ideal choice for social media data. In addition, HDBSCAN can deal well with clusters that vary in shape and size \cite{Campello.2013}.}

{Subsequently, we label the clusters using automatic labeling using LLAMA-3 model \emph{Llama-3-70b-chat-hf} \cite{Llama3.2024}. We generate five names per cluster and manually pick the most fitting one by analyzing representative posts. We hereby obtain concise summaries of the primary themes within each cluster, thus providing clear and contextually relevant labels.}

\subsection{Modeling past latent vulnerability (Step~2)}
\subsubsection{IPS reweighting}
Given the users $\mathcal{U}$, the hate speech posts $\mathcal{H}$, and the user-hate speech interactions $S_{uh}$, we apply IPS reweighting to alleviate selection bias and obtain debiased propensity scores $\pi$. The propensity score describes the probability that a user is exposed to hate speech.

\textbf{Assumption}: As in \cite{Joachims.2017}, we assume that users share posts when (a)~they have been exposed to the post, denoted as binary variable $E_{uh} \in \{0,1\}$ ($E_{uh} = 1$ if user $u$ was exposed to hate speech post $h$, $E_{uh} = 0$, otherwise), and (b)~when they are interested in the post, denoted as binary variable $I_{uh} \in \{0,1\}$ ($I_{uh} = 1$ if the user is interested in the post, $I_{uh} = 0$, otherwise). 
%This yields 
% \begin{equation}
%     S_{uh} = E_{uh} \cdot I_{uh}.
% \end{equation}
We further assume that resharing of hate speech is missing-not-at-random (MNAR) \cite{Little.2020}. Hence, the probability of user $u \in \mathcal{U}$ spreading a hate speech post $h \in \mathcal{H}$ is a product of the probability of exposure and the probability of interestingness of a post \cite{Saito.2020}. Formally, we have
\begin{equation}
\begin{split}
    P(S_{uh} = 1) & = \underbrace{P(E_{uh} = 1)}_{\text{exposure prob.}} \quad\cdot \underbrace{P(I_{uh} = 1)}_{\text{interestingness prob.}} \\
    & = \theta_{uh} \cdot \iota_{uh}  
\end{split}
\end{equation}
for all $S_{uh} \in S$ and where $\theta_{uh} > 0$ is the probability that user $u$ is exposed to hate speech post $h$ and $ \iota_{uh} > 0$ is the probability that user $u$ is interested in hate speech post $h$. Hence, the question arises how to compute $\theta_{uh}$. Our idea in the following is to leverage the propensity score $\pi$ for that, where the propensity score is defined as $\pi = P(E_{uh} = 1)$. However, the propensity score gives a biased estimate of $\theta_{uh}$ due to the observational nature of social media data, and, instead, we then use a debiased propensity score.

\textbf{Why the propensity score is a biased estimate of $\theta_{uh}$:}
Simply computing $\pi$ from our data gives a biased estimate for $\theta_{uh}$. The reason is that, when working with observational click data, we face two major challenges: \emph{First,} we only record positive interactions between users and hate speech posts (i.e., hate speech posts that are reshared by the user). Negative interactions (i.e., hate speech posts that were \emph{not} reshared by the users) are \emph{not} recorded. Therefore, we do \emph{not} know whether (a)~the user was not interested in a given post or whether (b)~the user was not exposed to the post. \emph{Second,} the observed interactions are likely subject to selection bias. Users are more likely to reshare hate speech posts that are more prevalent, and online platforms are more likely to recommend popular posts to users. Because of the above, the propensity scores $\pi$ is biased when computed from observational social media data, i.e., 
\begin{equation}
    \pi_{\mathrm{biased}} = P(S_{uh} = 1 \,\mid\, I_{uh} = 1) \neq \theta_{uh} 
    %=  P(E_{uh} = 1).
\end{equation}

\noindent
\textbf{Debiasing the propensity score:} Here, we propose to leverage IPS to learn debiased propensity scores based on the observed social media data. IPS acts as a reweighting mechanism that assigns larger weights to hate speech posts that are less likely to be observed, thereby creating a pseudo-randomized trial where users are shown posts at random. By doing so, we will later be able to learn debiased embeddings of the past latent vulnerability to hate speech. Following earlier research \cite{Cheng.2021}, we use three different estimation procedures to obtain unbiased propensity scores: 
\begin{enumerate}[leftmargin=*]
\item \emph{Virality-based propensity score:} This propensity score takes into account the virality of a hate speech post $h$, i.e.,
\begin{equation}
    \pi_h^{\mathrm{virality}} = \hat{\theta}^{\mathrm{virality}}_{h} = \left( \frac{\sum_{u\in \mathcal{U}} S_{uh}}{\max_{h \in \mathcal{H}} \sum_{u \in \mathcal{U}} S_{uh}} \right)^\mu ,
\end{equation}
where we assume that the probability that a user $u$ is exposed to a hate speech post $h$ is higher if the post has gone more viral (as measured by the number of reshares in the bipartite graph). The IPS for rare events could be very high. Hence, we apply a smoothing operator $\mu = 0.5$ for more stable propensity scores (see Appendix~\ref{sec:robustness_checks} for a robustness check with varying $\mu$).

\item \emph{Follower-based propensity score:} This propensity score takes into account the virality of the hate speech post $h$ as well as the popularity of the resharing user $u$ (as measured as the number of followers $F_u$). Formally, it is given by 
\begin{equation}
    \pi_h^{\mathrm{follower}} = \hat{\theta}^{\mathrm{follower}}_{uh} = \left( \frac{\sum_{u\in \mathcal{U}} S_{uh} \cdot F_u}{max_{h \in \mathcal{H}} \sum_{u \in \mathcal{U}} S_{uh} \cdot F_u} \right)^\mu,
\end{equation}
with $\mu = 0.5$. Here, we also account for the bias induced by the popularity of the user, since popular users are more likely to be exposed to hate speech \cite{ElSherief.2018}.

\item \emph{Neural network-based propensity score:}
This propensity score is estimated by a neural network via
\begin{equation}
    \pi_h^{\mathrm{neural}} = \hat{\theta}^{\mathrm{neural}}_{h} = \sigma({e_h}),
\end{equation}
where ${e_h}$ is the content of a given hate speech post as learned by LDA topic modeling and $\sigma()$ is the sigmoid function. This way, we implicitly encode the hate speech post's content in the latent representation.
\end{enumerate}
As a result, we obtain debiased propensity scores. Later, we perform an ablation study where we compare the three variants (1)--(3) of our recommendation algorithm to the biased algorithm and find that the debiased recommendation algorithms perform better.

\subsubsection{Modeling past latent vulnerability}

To model the past latent vulnerability (PLV) of users to hate speech (which gives our control variable ${X}$), we build upon concepts from recommendation systems \cite{Bonner.2018}. Thereby, we address one particular difficulty in that both the interestingness $I_{uh}$ and the exposure $E_{uh}$ are \emph{not} observable. Rather, only the interactions $S_{uh}$ between users and hate speech posts are observable on social media platforms. Hence, we learn a recommender algorithm that ranks content according to the interestingness of users simply using the observed interactions. 

\textbf{Recommender algorithm:} In our framework, we employ a Bayesian personalized ranking matrix factorization (BPRMF) \cite{Rendle.2009}. We optimize the parameters in the BPRMF using a pairwise Bayesian personalized ranking (BPR) loss \cite{Rendle.2009}. By using BPRMF, we address the challenges of missing negative interactions, since it assumes that the observed interactions better explain users' preferences than unobserved ones and should hence be assigned a higher predictive score. Since we employ three different ways to debias propensity scores, we also obtain three different variants of our debiased recommendation model: \textbf{BPRMF-V} (using $\pi_{\mathrm{virality}}$); \textbf{BPRMF-F} (using $\pi_{\mathrm{follower}}$); and \textbf{BPRMF-NN} (using $\pi_{\mathrm{neural}}$). In Appendix~\ref{sec:robustness_checks} we perform a robustness check where we compare the performance of our unbiased recommendation models and the biased recommender. We find that the recommendation algorithms with debiased propensity scores are better at predicting the users vulnerability to hate speech. For each user $u$, we model PLV to hate speech as a 64-dimensional embedding ${X}_u$. For comparison, the base model \textbf{BPRMF} is biased since it uses $\pi_{\mathrm{biased}}$ as the propensity score. 

\textbf{Loss:} Formally, let $\mathcal{D} = \mathcal{U} \times \mathcal{H} \times \mathcal{H}$ be the set of all observed interactions $(u,h)$ and unobserved interactions $(u,g$), for each pair $(h,g)$ of hate speech posts, where $h \neq g$. We model PLV as a personalized ranking task with a recommendation algorithm from observed interaction. Based on previous research \cite{Saito.2020}, the \emph{ideal loss of pairwise BPR} is:
\begin{equation}
    L_{\mathrm{ideal}}(\hat{R}) = \frac{1}{|\mathcal{D}|} \sum_{(u,h,g) \in \mathcal{D}} \iota_{uh} \large(1 - \iota_{ug} \large) \, \ell(\hat{R}_{uhg}),
\end{equation}
where $\ell = -\mathrm{ln} \sigma(\cdot)$ is the local loss for the triplet $(u,h,g)$, $\iota_{uh}$ is the interestingness of the post $h$ for user $u$, and $\hat{R}_{uhg}$ is the difference between the predicted scores for hate speech posts $h$ and $g$. However, the \emph{interestingness is unobservable} and only the interactions $S_{uh}$ are available. BPR then assumes that \emph{interacted messages should be ranked higher} than all the other non-interacted messages and optimizes the following loss function to obtain latent factors:
\begin{equation}
    L_{\mathrm{BPR}}(\hat{R}) = \frac{1}{|\mathcal{D}|} \sum_{(u,h,g) \in \mathcal{D}} S_{uh} \large(1 - S_{ug} \large) \, \ell(\hat{R}_{uhg}).
\end{equation}
However, the loss function of the BPR model is biased towards the ideal pairwise loss, because BPR treats all non-interacted messages as if the user was not interested in them and does not deal with messages the user was not exposed to \cite{Saito.2020}. To deal with the \emph{missing data in implicit feedback}, we follow previous research \cite{Cheng.2021} and employ propensity scores as $\theta_{uh}$ for the unbiased estimator of the ideal pairwise loss as follows
\begin{equation}
    L_{\mathrm{unbiased}}(\hat{R}) = \frac{1}{|\mathcal{D}|} \sum_{(u,h,g) \in \mathcal{D}} \frac{S_{uh}}{\theta_{uh}} \large(1 - \frac{S_{ug}}{\theta_{ug}} \large) \, \ell(\hat{R}_{uhg}).
\end{equation}
Of note, propensity scores can often become extremely small, e.g., hate speech posts with low virality also have low exposure probability. This leads to large variances in IPS-based approaches. We reduce the variance by employing a non-negative loss \cite{Saito.2020} given by
\begin{equation}
    \hat{L}_{\textrm{non-neg}}(\hat{R}) = \frac{1}{|\mathcal{D}|} \sum_{(u,h,g) \in \mathcal{D}} \max\{  \ell(\hat{R}_{uhg}), 0 \}.
\end{equation}
In addition, we add a $l_2$-regularization for the user embeddings and post embeddings, ${U}$ and ${H}$, respectively. The \textbf{final loss function} is
\begin{equation}
    \argmin_{{U}, {H}}\hat{L}_{\textrm{non-neg}}(\hat{R}) + \lambda \large( ||{U}||_2^2 + ||{H}||_2^2 \large),
\end{equation}
where $\lambda$ is the weight of the $l_2$-regularization for the latent representations.

\subsection{Modeling hate speech resharing (Step~3)}
\subsubsection{Likelihood to reshare hate speech}
{For each user $u$, we now compute the probability to reshare hate speech in general $Y_u$ as well as the probability to reshare different types of hate speech $Y_{uc}$. Formally, we define the user's probability to reshare hate speech $Y_u \in [0,1]$  as
\begin{equation}
    Y_u = \frac{n^{u}_{\mathrm{hate}} }{n^{u}_{\mathrm{hate}} + n^{u}_{\mathrm{normal}}} ,
\end{equation}
where $n^{u}_{\mathrm{hate}}$ and $n^{u}_{\mathrm{normal}}$ are the number of hate speech posts and normal posts that user $u$ shared, respectively. }

{Similarly, the probability to reshare a specific type of hate speech $Y_{uc} \in [0,1]$ is defined as
\begin{equation}
    Y_{uc} = \frac{n^{u}_c }{n^{u}_{\mathrm{hate}} + n^{u}_{\mathrm{normal}}} ,
\end{equation}
where $n^{u}_c$ is the number of hate speech posts that user $u$ has shared from cluster $c$.}

\subsubsection{Effect estimation}
We now model the relationships between the user attributes $A$ and the users' probability to reshare hate speech $Y$ while controlling for PLV ${X}$ for each user $u$. We model the effects using an explainable boosting machine~(EBM) \cite{EBM}. EBM is a tree-based, cyclic gradient boosting GAM (generalized additive model) with automatic interaction detection. It provides state-of-the-art performance while being fully explainable.

In our work, we are interested in understanding the effects of the following user attributes $a \in A$: verified ($= 1$ if the user is verified, $0$ otherwise), account age (in days), \#posts (number of posts the user did), \#followers, and \#friends. Formally, we model
\begin{equation}
    g(E[Y_u]) = \beta_0 + \sum f_m(a_m) + \sum f({X}_u),
\end{equation} 
where $g$ is the link function that adapts the GAM to the regression task. For simplicity, we use $Y_u$ in above equation, which can be replaced by $Y_{uc}$ for each hate speech cluster $c \in C$. By using EBM instead of traditional, linear regression \cite{Cheng.2021}, we are able to model the non-linear effects of each user attribute on the users' probability to reshare hate speech. Still, our results are \emph{explainable} in that we can interpret the estimated effects. 

%Under certain assumptions, we can interpret our findings as causal effects. In Appendix~\ref{sec:establishing_causality}, we elaborate on how to establish causality, show our causal graph, discuss the validity of assumptions, perform a causal sensitivity analysis to account for the possibility of unobserved confounding, and evaluate the performance with semi-synthetic data.

% For each user $u$, we have a vector ${a}_u = (a_1, \ldots, a_m)$, with ${a} \in A$, that encodes the user's attributes, the PLV $X_u$ and the probability to reshare hate speech $Y_u$ ($Y_{uc}$ for different types of hate speech $c$).

% \textbf{Explainable boosting machine:}
% In our framework, we model the effects of user attributes $A$ on users' probability to reshare hate speech $Y$, while controlling for PLV $X$, with an explainable boosting machine~(EBM) \cite{EBM}. EBM is a tree-based, cyclic gradient boosting GAM (generalized additive model) with automatic interaction detection. It provides state-of-the-art performance while being fully explainable. 

\section{Experimental Setup}

\subsection{Data}

To understand who shares hate speech and which type, we apply our framework to a comprehensive, state-of-the-art dataset \cite{Founta.2018}. The dataset consists of normal and hate speech posts from the social media platform X (formerly Twitter) that were human-labeled following the definitions and procedures described in \cite{Founta.2018}. For all posts in the original dataset, we then additionally collected the corresponding reshares. As a result, our augmented dataset consists of 25,196 root posts (96.05\% normal and 3.95\% hateful posts) from 31,905 users. The root posts received $N =$ 65,920 reshares by 61,961 users, out of which 43,929 (66.64\%)  were reshares of normal root posts and 21,991 (33.36\%) were reshares of hateful root posts. As such, hateful root posts were reshared disproportionally often.

\textbf{Model variables:} We analyze the effects of five user attributes on the probability to reshare: $\boldsymbol{a}_1 $: verified ($=1$ if the user is verified, 0 otherwise); $\boldsymbol{a}_2$: account age (in days); $\boldsymbol{a}_3$: \#post (number of posts by the user); $\boldsymbol{a}_4$: \#follower (number of followers of the user); and $\boldsymbol{a}_5$: \#friend (number of friends of the user). The distributions of \#post, \#follower, and \#friend are highly right-skewed, because of which we log these user attributes. Appendix Table~\ref{tab:descriptives_user_attributes} shows summary statistics for the different user attributes.

% 24,201 normal root posts and 995 hateful root posts
% 43,929 retweets of normal root posts
% 21,991 retweets of hateful root posts

\textbf{Ethical statement:} Data collection was conducted by following standards for ethical research \cite{Rivers.2014}. Our analyses are based on publicly available data, and we only report aggregated results. We respect users' privacy by not publishing usernames and do not attempt to track or de-identify users who decided to remain anonymous. We have received ethical approval by our ethics committee (=IRB) at the University of (anonymized). The approval number is (anonymized).
%EK-MIS-2023-160.

\subsection{Baselines}

In this work, we leverage a debiasing framework to model the effects of user attributes on users' probability to reshare hate speech using an EBM. We refer to our framework as \textbf{DF-EBM}. Note that, for benchmarking, our setting requires baselines that provide explainable results since we aim to provide an understanding of the user attributes that make users reshare hate speech. To the best of our knowledge, we are aware of only one benchmark where the effects of user attributes on the probability to reshare content were modeled: (1)~\textbf{DF-linear} \cite{Cheng.2021} performs debiasing but, in contrast to our framework, is fully \emph{linear}. Formally, instead of allowing for non-linear relationships in step~\circledorange{3}, \textbf{DF-linear} models linear effects of user attributes on the probability to reshare content. It is thus explainable as it outputs coefficients for each user attribute, but it cannot capture non-linearities and is thus \emph{not} flexible since the modeled effects are assumed to be linear. 

In addition, we design the following variants of our debiasing framework that we also use as baselines: (2)~\textbf{DF-ExNN} where we replace the EBM with an explainable neural network \cite{Yang.2020}. The explainable neural network enhances the explainability of a vanilla neural network by adding architecture constraints, which is to achieve a balance between prediction performance and model explainability. (3)~\textbf{DF-NAM} where we replace the EBM with a neural additive model \cite{Agarwal.2021}, which learns a linear combination of neural networks that each attend to a single input feature. This way, it remains explainable while being able to model non-linear effects.

% \subsection{Performance metrics}

% For evaluation, we split our dataset into train/test set (80/20). Our performance metrics are two-fold. (1)~To evaluate the performance of the entire framework (step~\circledorange{3}), we use the root mean squared error (RMSE). (2)~We additionally evaluate the performance of the recommendation via recall@$k$ and NDCG@$k$. The former is a measure of the posts that were interesting to the user out of those that the model predicted to be interesting as the top-$k$ posts. The latter measures the accuracy of the recommendation by comparing the ground-truth interestingness to the predicted ranking of posts for the top-$k$ ranked posts.

\subsection{Implementation details}

We implemented our framework in Python~3.8. {For the hate speech clustering, we use BERTopic model \emph{all-MiniLM-L6-v2} \cite{Grootendorst.2022.bertopic} to generate embeddings and UMAP to reduce the embeddings into a seven-dimensional space. To cluster, we use HDBSCAN with a minimum cluster size of 10. We name the resulting cluster with the LLAMA-3 model \emph{Llama-3-70b-chat-hf} \cite{Llama3.2024}.} To model the effects, we use the EBMRegressor of InterpretML \cite{EBM}. We allow for ten interactions in the EBM model and run it for a maximum of 5,000 rounds. Moreover, we performed hyperparameter tuning (see Appendix~\ref{sec:hyperparam}). For evaluation, we split our dataset into train/test set (80/20). To evaluate the performance of the entire framework (step~\circledorange{3}), we use the root mean squared error (RMSE). We average the performance of the EBMRegressor as well as our baselines over five runs. To model PLV, we use BPRMF \cite{Rendle.2009}. We set the learning rate of our BPRMF model to 0.001, the batch size to 64, and the embedding size to 64. We average the performance over five runs with varying train/test splits. For the IPS reweighting with $\pi^{neural}_h$, we employ linear discriminant analysis (LDA) topic modeling\footnote{\url{https://scikit-learn.org/stable/modules/generated/sklearn.discriminant_analysis.LinearDiscriminantAnalysis.html}} to get a latent representation $e_h$ of the topics in the posts.

\section{Results}
% In this work, we predict the probability of users to reshare hate speech based on their user attributes while controlling for confounding due to the latent vulnerability of users. 

\subsection{Hate speech cluster}
{We begin by reporting the hate speech clusters resulting from our topic modeling. We cluster the 270 hate speech root posts into four clusters, which we labeled automatically. The optimal number of clusters is determined automatically by HDBSCAN, which allows us more flexibility since we do not have to determine the number of clusters beforehand \cite{Campello.2013}.}

{The largest cluster contains 123 posts and got the label \say{racist and misogynistic rants}. The posts in this cluster received a total of 15.219 reshares and mostly contained racial and misogynistic slurs as well as angry, abusive, and offensive language. The second cluster contains 93 posts with \say{anti-Trump and anti-right wing rants} and received 6.207 reshares. The posts share political hate, specifically discriminating against Trump and right-wing politics. The third cluster contains 36 hate speech posts with the label \say{toxic rants} and received 512 reshares. These posts are hateful and offensive in general and do not have a common group that is being targeted. The last cluster contains only 16 posts with the label \say{social media noise}. The posts in this cluster are mostly spam and received 51 reshares. Together, this shows that most of the reshares in our dataset are racial and misogynistic hate, hate against Trump and right-wing politicians, as well as general hateful comments. Appendix~\ref{sec:app_clusters} shows the characteristic words and example posts for each cluster. }

% Performance for BPRMF-H data
\begin{table}[ht]
    \centering
    \footnotesize
    \begin{tabular}{lccc}
    \toprule
         Model& Explainable & Flexible & RMSE\\
    \midrule
         DF-linear & \includegraphics[width=2mm, height = 2mm]{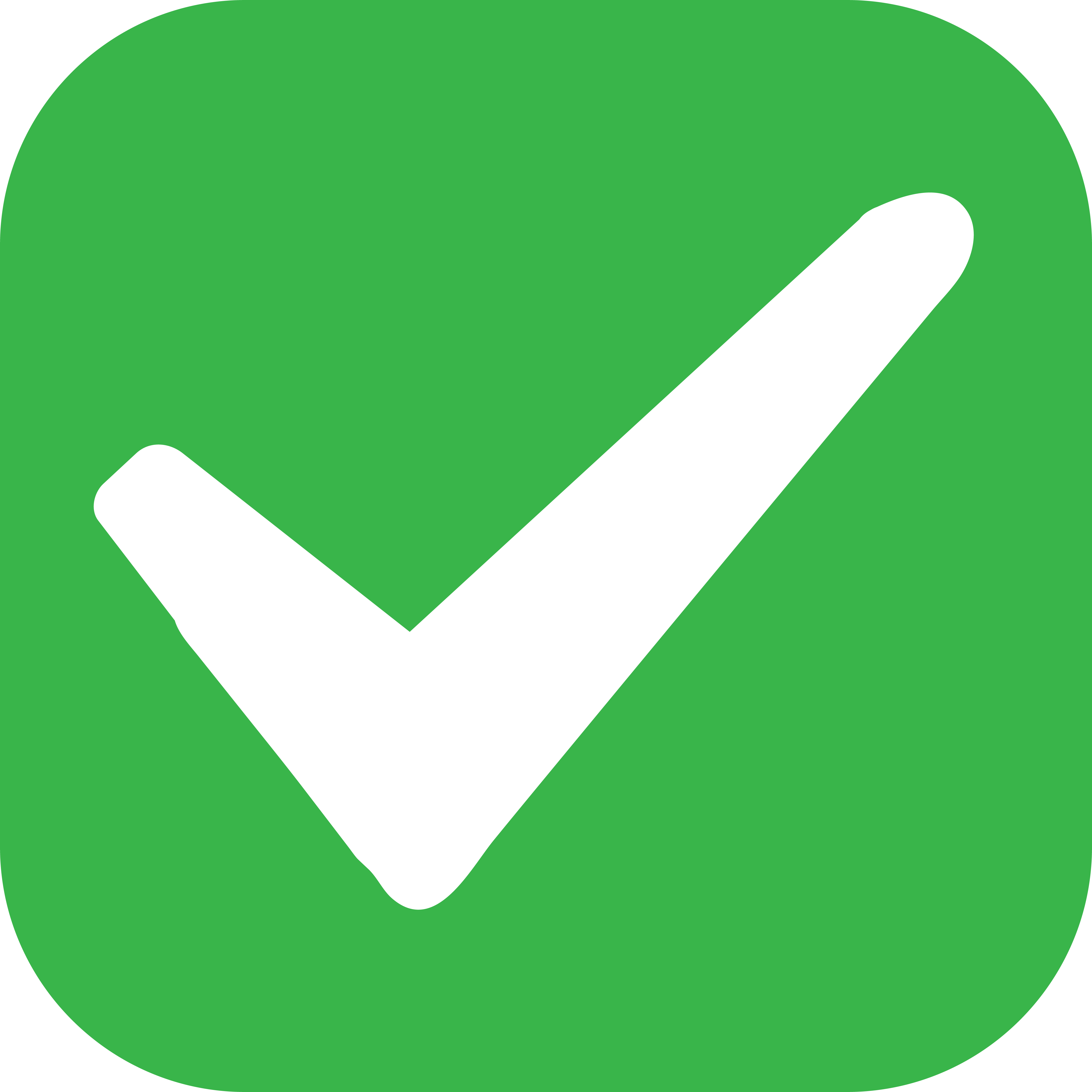} & \includegraphics[width=2mm, height = 2mm]{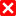} & 0.0885\\
         DF-ExNN  & \includegraphics[width=2mm, height = 2mm]{figures/tick-icon.png} & \includegraphics[width=2mm, height = 2mm]{figures/tick-icon.png} & 0.0911 \\
         DF-NAM & \includegraphics[width=2mm, height = 2mm]{figures/tick-icon.png} & \includegraphics[width=2mm, height = 2mm]{figures/tick-icon.png} &  0.1342 \\
         % MLP Regressor & No & 0.05 \\
         \midrule
         &  &  & \textbf{0.0863}\\
        \multirow{-2}{*}{\textbf{DF-EBM (ours)}} & \multirow{-2}{*}{\includegraphics[width=2mm, height = 2mm]{figures/tick-icon.png}} & \multirow{-2}{*}{\includegraphics[width=2mm, height = 2mm]{figures/tick-icon.png} } & \scriptsize($-$2.5\%) \\
    \bottomrule
    \end{tabular}
    \caption{Performance comparison (measured by the RMSE) for predicting users' probability to reshare hate speech. Our debiasing framework performs best (relative improvement of 2.5\% over the best baseline, whereby DF-linear is the only baseline that we are aware of in the literature).}
    \vspace{-0.4cm}
    \label{tab:performance_copmarison}
\end{table}

\subsection{Performance comparison}
We next report the performance of our framework when predicting the effects of user attributes on the users' probability to reshare hate speech. We compare our framework against multiple baselines based on the RMSE (root mean squared error). To the best of our knowledge, there is only one baseline in the literature: \textbf{DF-linear} \cite{Cheng.2021}, which builds upon a linear approach and thus can \emph{not} account for non-linearites. 
All of the baselines as well as our model are \emph{explainable} since we aim to provide an understanding of the user attributes that make users reshare hate speech.

Table~\ref{tab:performance_copmarison} shows the RMSE for each variant of our framework. Our debiasing framework \textbf{DF-EBM} outperforms the \textbf{DF-linear} baseline with a relative improvement of 2.5\%. The improvement is statistically significant according to Welch's $t$-test \cite{Welch.1947} at a $p < 0.001$ level. Importantly, both frameworks have access to identical data so the performance gain must be attributed solely to the flexibility of our framework in capturing nonlinearities. Moreover, our \textbf{DF-EBM} also outperforms the other custom baselines that are non-linear. Here, the Welch's $t$-tests \cite{Welch.1947} are again statistically significant at the $p < 0.01$ level. Together, this demonstrates that our debasing framework \textbf{DF-EBM} is superior.

\subsection{Analyzing hate speech resharing}

\begin{figure}[ht]
    \centering
    \includegraphics[width=0.6\linewidth]{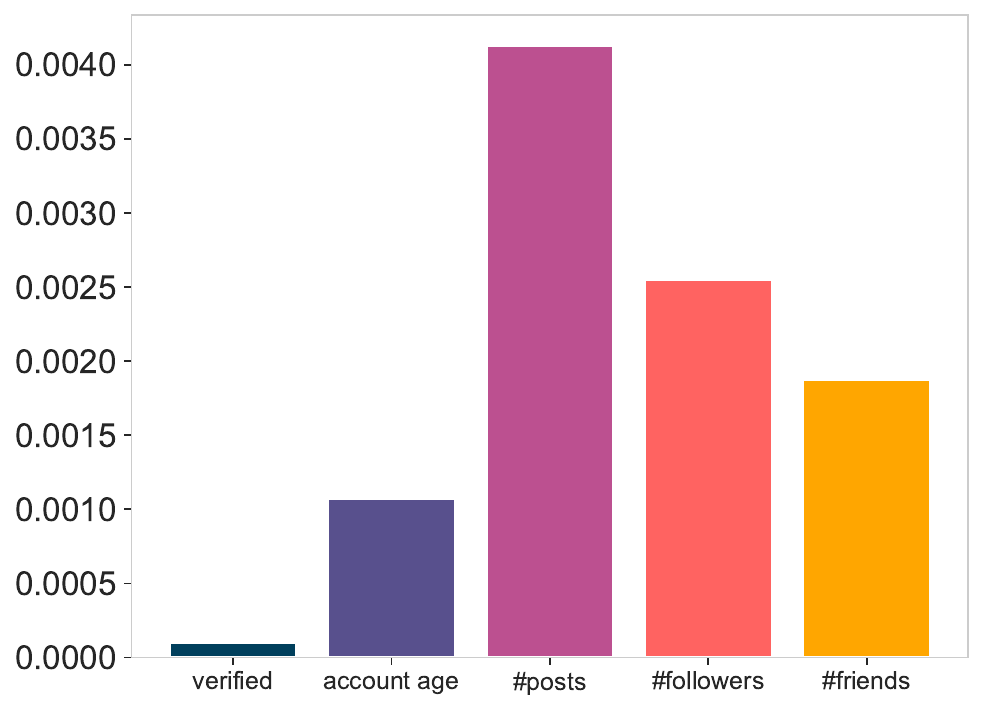}
    \vspace{-0.3cm}
    \caption{Feature importance of the user attributes when predicting the users' probability to reshare hate speech.}
    \label{fig:ebm_feature_importance}
\end{figure}

We now leverage the explainability of our framework to understand which user attributes affect the users' probability to reshare hate speech. For this, we analyzed the effects of users' verified status, account age, number of posts, number of followers, and number of friends. We first look at the feature importance for each user attribute, which is computed as the average absolute predicted value of each feature for the training set (i.e., for each feature, we predict the absolute value for each sample in the training set and then average these scores). This is weighted by the relative frequency of each feature in the train set. Here, we find that the number of posts is the most influential user attribute when predicting users' probability to reshare hate speech (see Figure~\ref{fig:ebm_feature_importance}). Moreover, the second most influential attribute is the number of followers. Thereby, we add to the literature where the popularity of a user and their activity level are important indicators as to whether they \emph{post} hate speech \cite{Hua.2020, Ribeiro.2018}, whereas we show that the number of posts and followers also cause a larger propensity to \emph{reshare} hate speech.

\begin{figure*}[ht]
    \centering
    \begin{subfigure}[c]{0.24\textwidth}
        \centering
        \includegraphics[width=\textwidth]{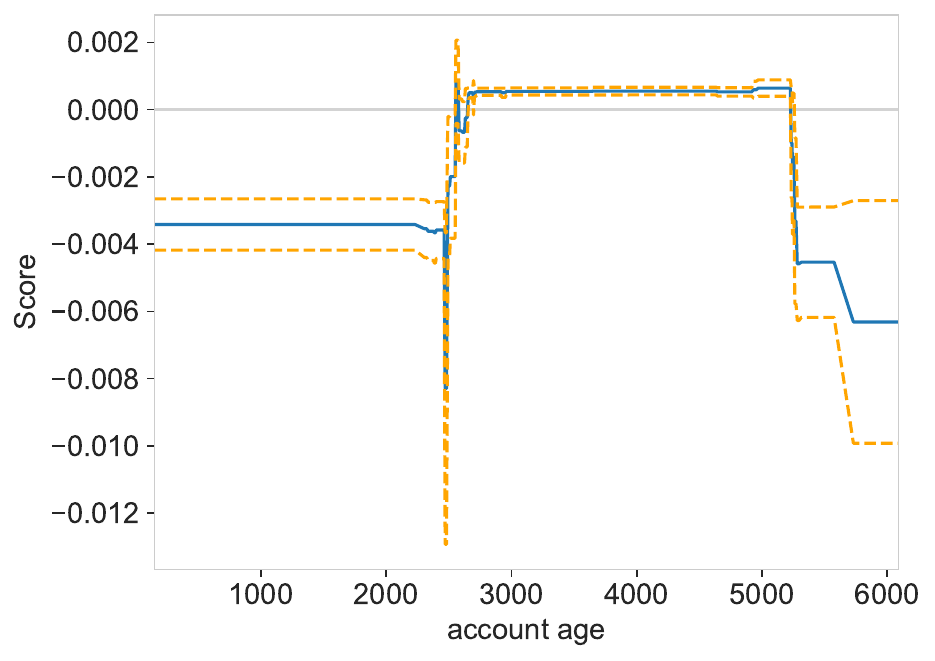}
        \caption{account age}
        \label{fig:score_account_age}
    \end{subfigure}
    ~
    \begin{subfigure}[c]{0.24\textwidth}
        \centering
        \includegraphics[width=\textwidth]{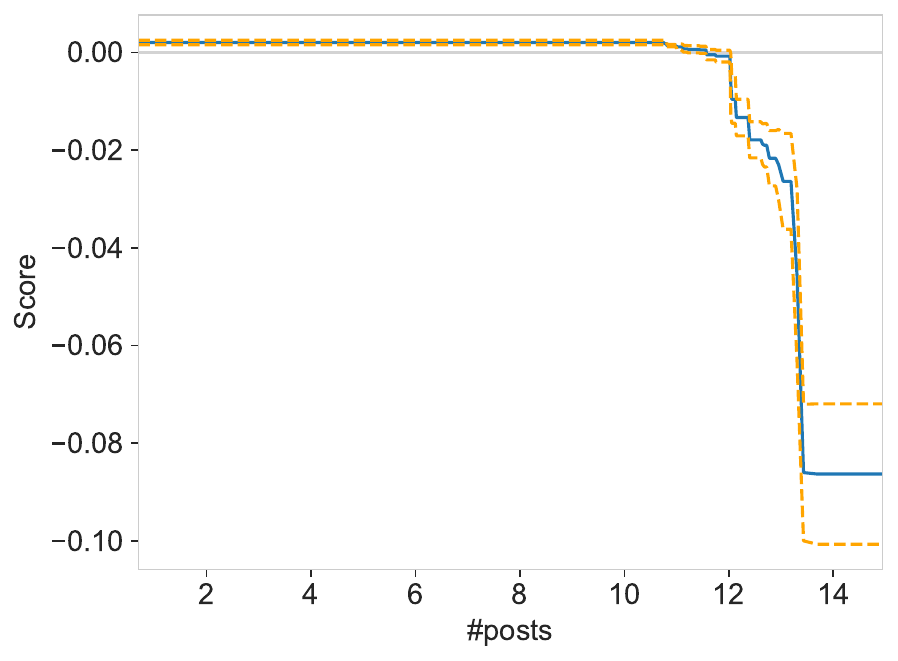}
        \caption{\#posts (in log scale)}
        \label{fig:score_posts}
    \end{subfigure}
    ~
    \begin{subfigure}[c]{0.24\textwidth}
        \centering
        \includegraphics[width=\textwidth]{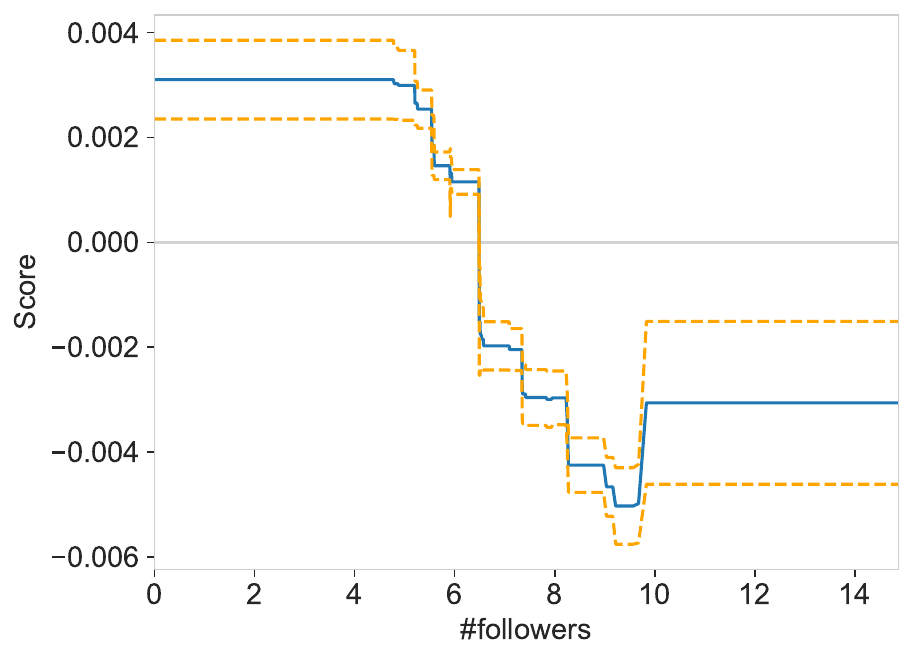}
        \caption{\#followers (in log scale)}
        \label{fig:score_followers}
    \end{subfigure}
    ~
    \begin{subfigure}[c]{0.24\textwidth}
        \centering
        \includegraphics[width=\textwidth]{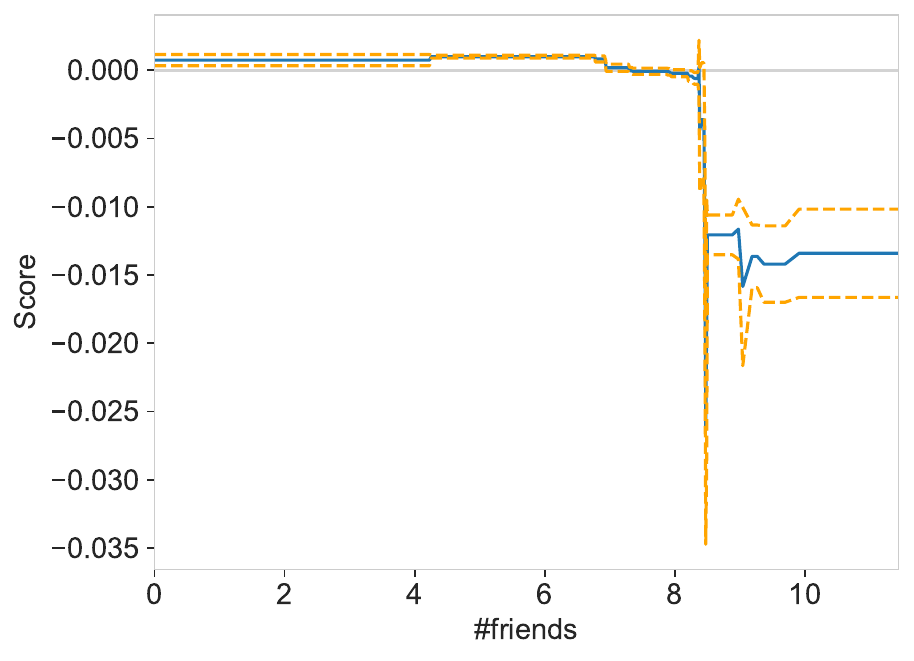}
        \caption{\#friends (in log scale)}
        \label{fig:score_friends}
    \end{subfigure}
    \vspace{-0.3cm}
    \caption{Contributions of the user attributes to the users' probability to reshare hate speech as estimated by our debiasing framework (both y-axis and outcome $Y$ are shown on the same scale). Each graph is normalized for better comparability such that the average prediction on the train set is at 0. The blue curve shows the predicted effect, while the orange curves show the upper and lower bounds of our model.}
    \vspace{-0.3cm}
    \label{fig:causal_effects}
\end{figure*}

We next look at the effects of each user attribute on the probability to reshare hate speech. Figure~\ref{fig:causal_effects} shows the contribution of each user attribute. The contribution is shown on the same scale as the output variable, and the graph is normalized so that the average prediction is at 0. For example, a score of $-0.1$ at an account age of 100 days would mean that a user with a 100-day-old account is 10 percentage points less likely to reshare hate speech than the average prediction. Overall, we find that users with fewer posts, fewer followers, and fewer friends are more likely to reshare hate speech. In return, newer accounts are less likely to do so. Moreover, we find that verified users are 0.016 ($\pm 0.004$) percentage points more likely to reshare hate speech than the average prediction, which is contrary to our expectations that a verified status is typically linked to users with a large reputation who should thus engage less with hate speech. We further perform a comparison of PLV to hate speech and PLV to normal content (see Appendix~\ref{sec:comparison_normal_content} for the complete analysis).

% The average contribution for each user attribute is as follows: verified $=0.01329$; account age $=-0.000014$; \#posts $=-0.00041$; \#followers $=-0.000057$; and \#friends$=-0.000043$. 

\subsection{Resharing of types of hate speech}
{We now split the hate speech posts into their respective types and analyze the resharing for each type separately. Here, we look at the effects of users' verified status, account age, number of posts, followers, and friends on the resharing behavior. Similar to above, the contribution of each user attribute is shown on the same scale as the output variable. }

\begin{figure*}[ht]
    \centering
    \begin{subfigure}[c]{0.24\textwidth}
        \centering
        \includegraphics[width=\textwidth]{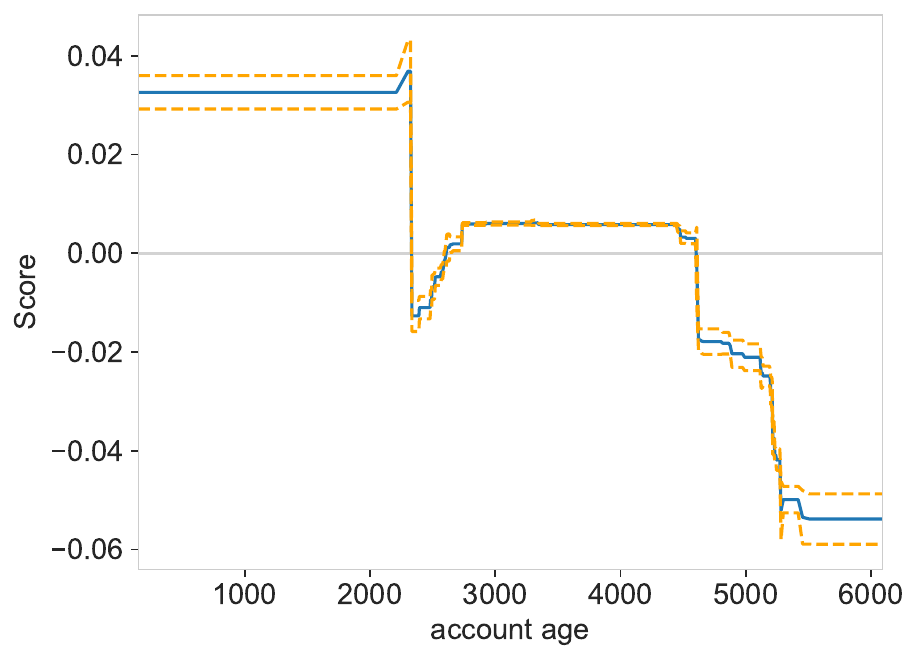}
        \caption{account age}
        \label{fig:score_account_age_racist}
    \end{subfigure}
    ~
    \begin{subfigure}[c]{0.24\textwidth}
        \centering
        \includegraphics[width=\textwidth]{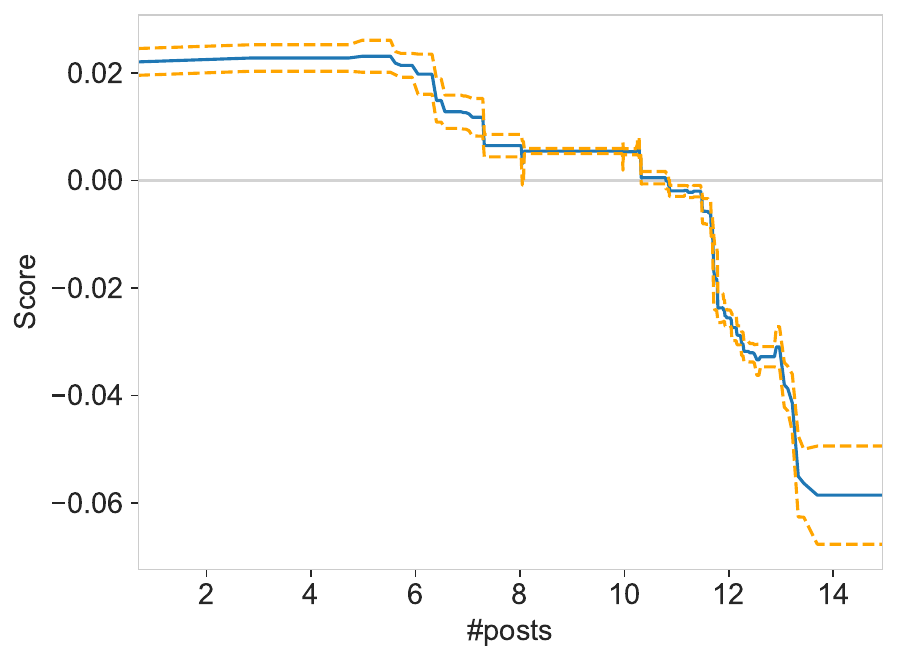}
        \caption{\#posts (in log scale)}
        \label{fig:score_posts_racist}
    \end{subfigure}
    ~
    \begin{subfigure}[c]{0.24\textwidth}
        \centering
        \includegraphics[width=\textwidth]{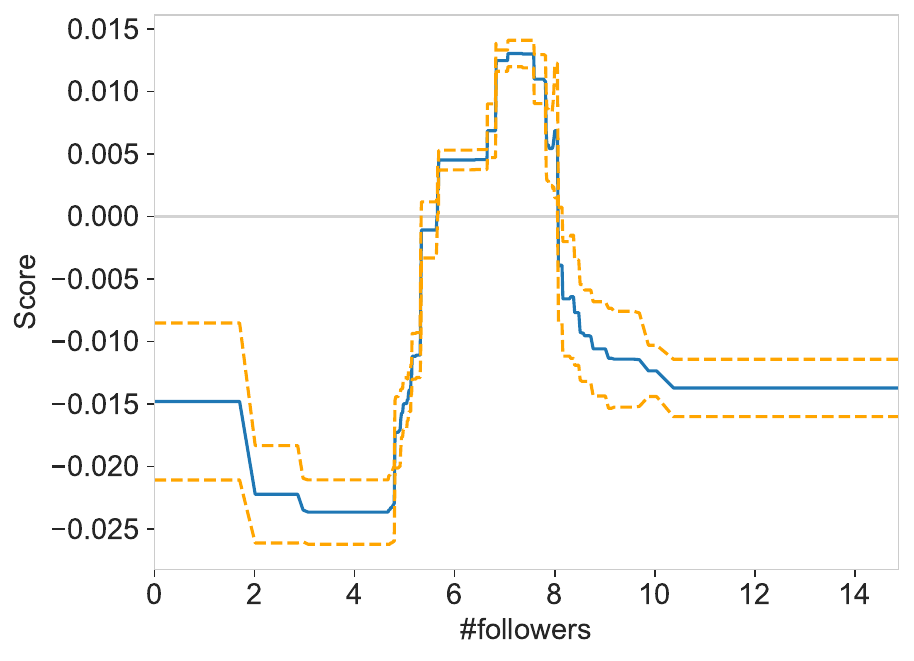}
        \caption{\#followers (in log scale)}
        \label{fig:score_followers_racist}
    \end{subfigure}
    ~
    \begin{subfigure}[c]{0.24\textwidth}
        \centering
        \includegraphics[width=\textwidth]{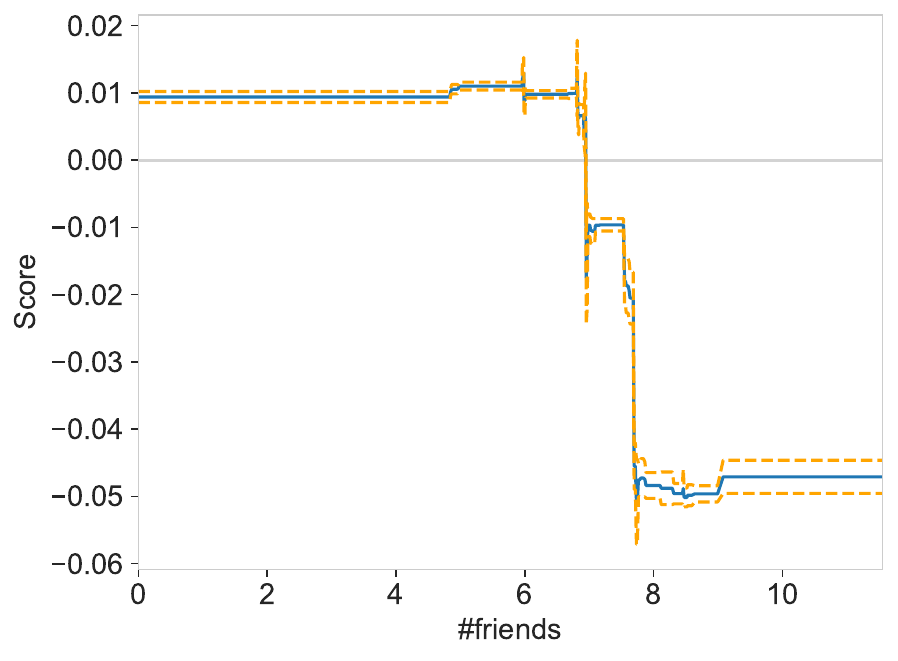}
        \caption{\#friends (in log scale)}
        \label{fig:score_friends_racist}
    \end{subfigure}
    \vspace{-0.3cm}
    \caption{Contributions of the user attributes to the users' probability to reshare racist and misogynistic rants as estimated by our debiasing framework (both y-axis and outcome $Y$ are shown on the same scale). Each graph is normalized for better comparability such that the average prediction on the train set is at 0. The blue curve shows the predicted effect, while the orange curves show the upper and lower bounds of our model.}
    \vspace{-0.3cm}
    \label{fig:causal_effects_racist}
\end{figure*}

{For \say{racial and misogynistic rants}, we find that newer accounts, accounts with fewer posts, and accounts with fewer friends tend to share more racist and misogynistic rants (see Figure~\ref{fig:causal_effects_racist}). At the same time, users with very few or a lot of followers share less racist and misogynistic rants. Verified users are 0.083 ($\pm 0.0002$) percentage points less likely to reshare racist and misogynistic rants. These findings are in part similar to overall hate sharing, where users with fewer posts and friends also tend to share more hate in general. }

\begin{figure*}[ht]
    \centering
    \begin{subfigure}[c]{0.24\textwidth}
        \centering
        \includegraphics[width=\textwidth]{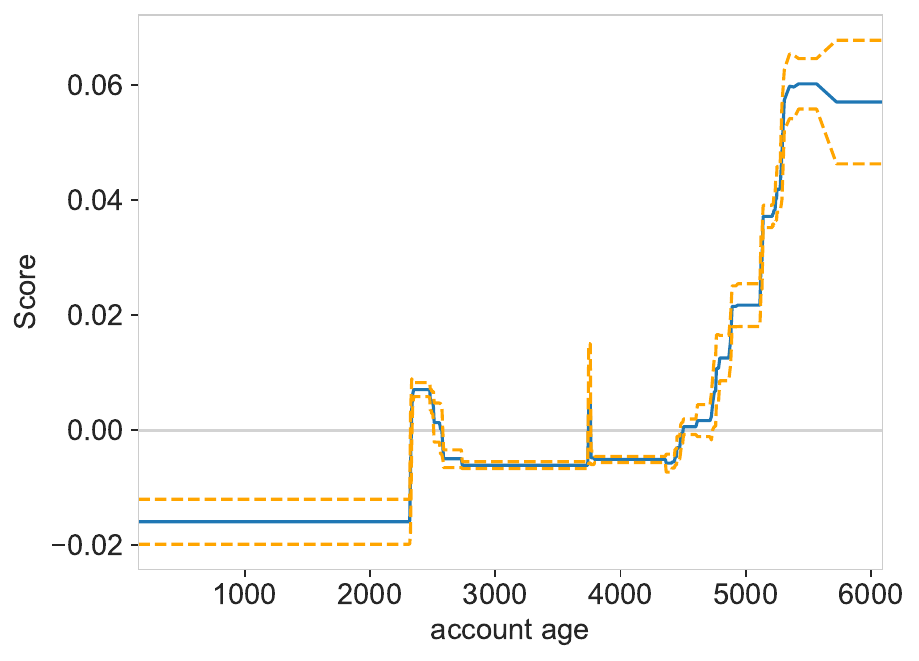}
        \caption{account age}
        \label{fig:score_account_age_political}
    \end{subfigure}
    ~
    \begin{subfigure}[c]{0.24\textwidth}
        \centering
        \includegraphics[width=\textwidth]{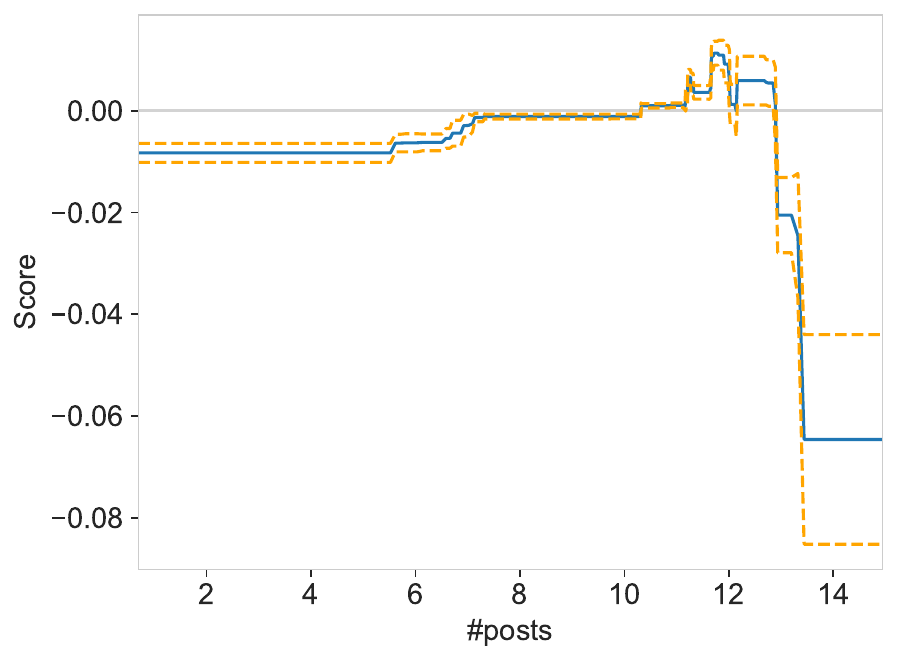}
        \caption{\#posts (in log scale)}
        \label{fig:score_posts_political}
    \end{subfigure}
    ~
    \begin{subfigure}[c]{0.24\textwidth}
        \centering
        \includegraphics[width=\textwidth]{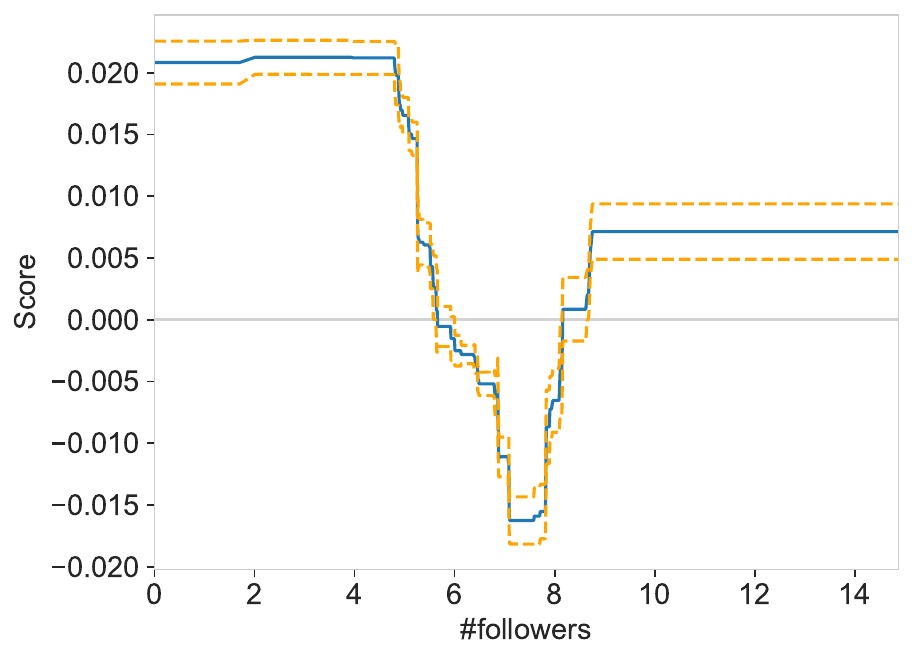}
        \caption{\#followers (in log scale)}
        \label{fig:score_followers_political}
    \end{subfigure}
    ~
    \begin{subfigure}[c]{0.24\textwidth}
        \centering
        \includegraphics[width=\textwidth]{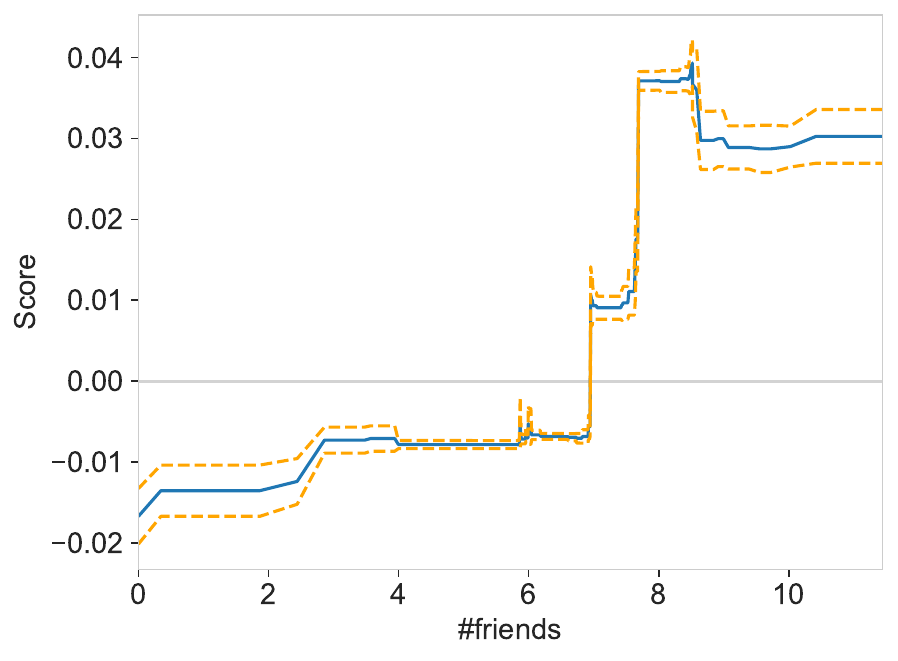}
        \caption{\#friends (in log scale)}
        \label{fig:score_friends_political}
    \end{subfigure}
    \vspace{-0.3cm}
    \caption{Contributions of the user attributes to the users' probability to reshare anti-Trump and anti-right wing rants as estimated by our debiasing framework (both y-axis and outcome $Y$ are shown on the same scale). Each graph is normalized for better comparability such that the average prediction on the train set is at 0. The blue curve shows the predicted effect, while the orange curves show the upper and lower bounds of our model.}
    \vspace{-0.3cm}
    \label{fig:causal_effects_political}
\end{figure*}

{When looking at \say{anti-Trump and anti-right wing rants}, we find opposite effects to racial and misogynistic rants: newer accounts, users with fewer posts, and users with fewer friends share less anti-Trump hate (see Figure~\ref{fig:causal_effects_political}). Also the contribution curve for the number of followers is reversed, where now users with very few or a lot of followers share less anti-Trump hate. Moreover, the effect of verified status is reversed with verified users being 0.075 ($\pm 0.0002$) percentage points more likely to reshare anti-Trump hate. }

\begin{figure*}[ht]
    \centering
    \begin{subfigure}[c]{0.24\textwidth}
        \centering
        \includegraphics[width=\textwidth]{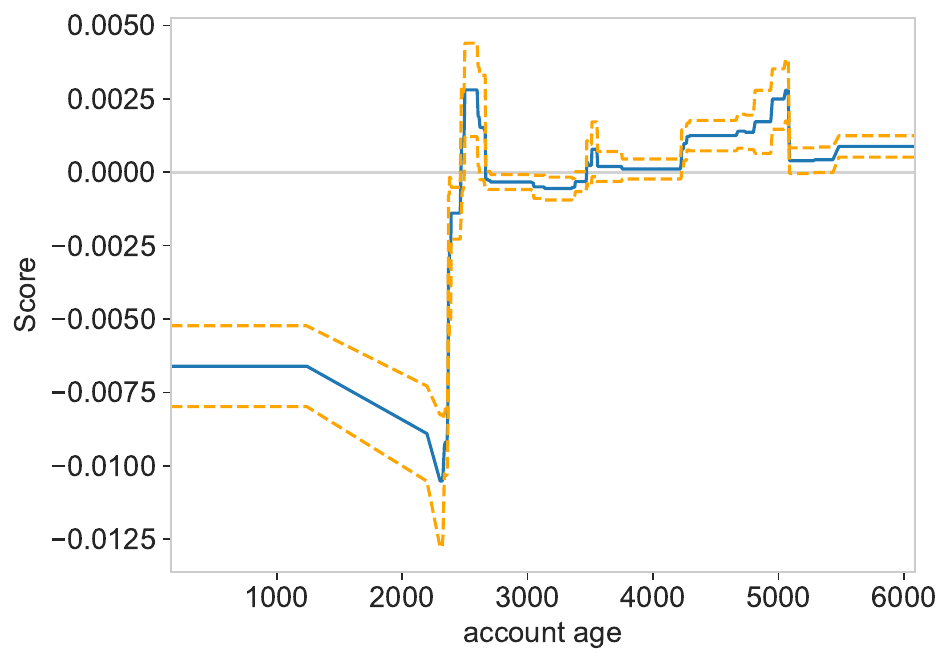}
        \caption{account age}
        \label{fig:score_account_age_toxic}
    \end{subfigure}
    ~
    \begin{subfigure}[c]{0.24\textwidth}
        \centering
        \includegraphics[width=\textwidth]{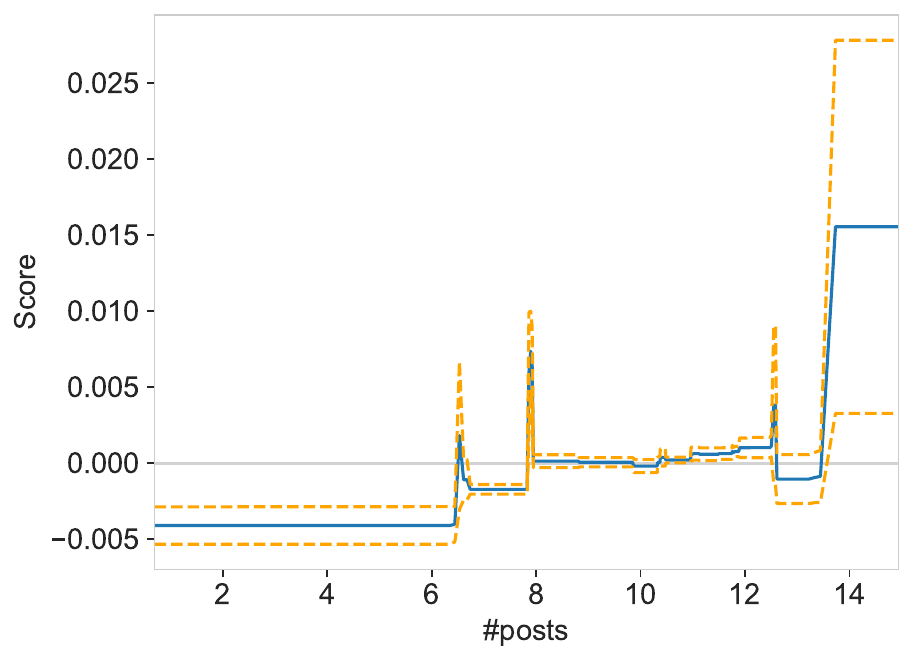}
        \caption{\#posts (in log scale)}
        \label{fig:score_posts_toxic}
    \end{subfigure}
    ~
    \begin{subfigure}[c]{0.24\textwidth}
        \centering
        \includegraphics[width=\textwidth]{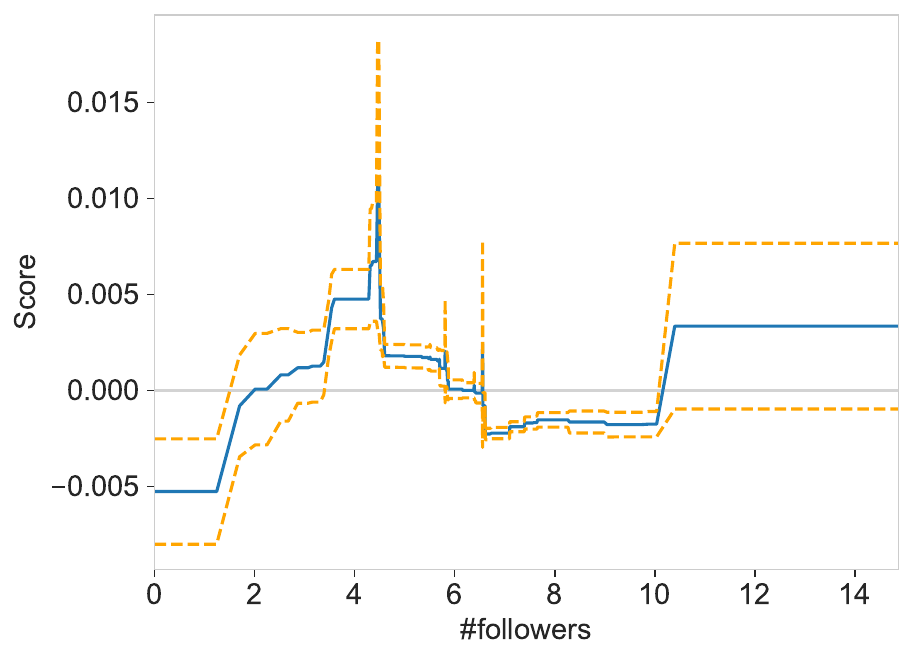}
        \caption{\#followers (in log scale)}
        \label{fig:score_followers_toxic}
    \end{subfigure}
    ~
    \begin{subfigure}[c]{0.24\textwidth}
        \centering
        \includegraphics[width=\textwidth]{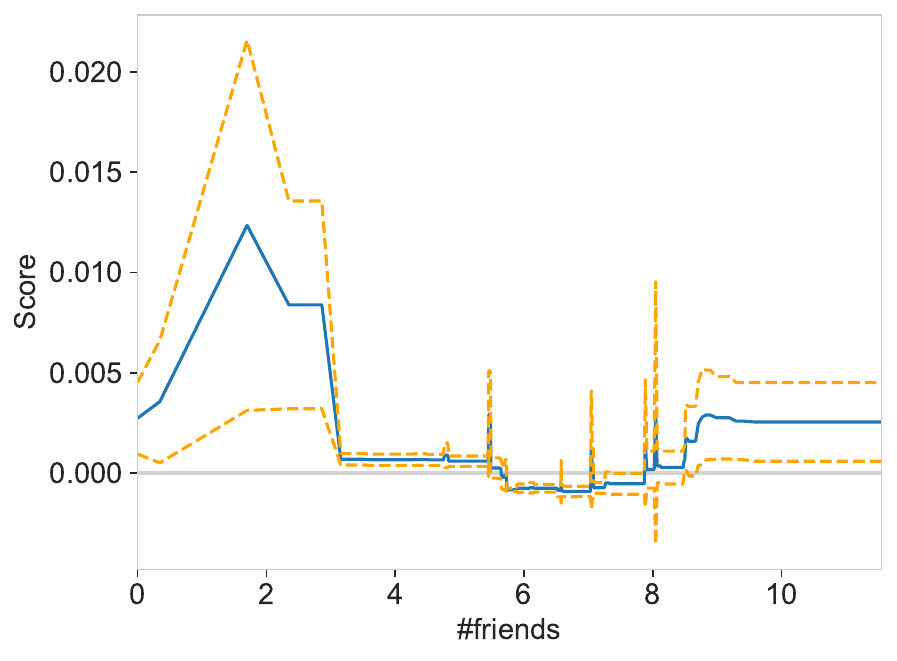}
        \caption{\#friends (in log scale)}
        \label{fig:score_friends_toxic}
    \end{subfigure}
    \vspace{-0.3cm}
    \caption{Contributions of the user attributes to the users' probability to reshare toxic rants as estimated by our debiasing framework (both y-axis and outcome $Y$ are shown on the same scale). Each graph is normalized for better comparability such that the average prediction on the train set is at 0. The blue curve shows the predicted effect, while the orange curves show the upper and lower bounds of our model.}
    \vspace{-0.3cm}
    \label{fig:causal_effects_toxic}
\end{figure*}

{For \say{toxic rants}, we find similar trends to anti-Trump hate resharing: younger accounts and users with fewer posts also share less toxic rants (see Figure~\ref{fig:causal_effects_toxic}). The effects for the number of followers and friends are different in that users with fewer followers share less toxic rants and users with very few or a lot of friends share more toxic rants. Verified users are 0.001 ($\pm 0.000$) percentage points less likely to reshare toxic rants.}

\subsection{Ablation study}
\label{sec:ablation_study}
We performed an ablation study to analyze the differences in performance when using different inputs to the past latent vulnerability (PLV). In particular, we compare four different inputs: (1) a base model where we do not control for PLV; (2) BPRMF-V model with the virality-based PLV using $P_{\mathrm{virality}}$; (3) BPRMF-F with the follower-based PLV using $P_{\mathrm{follower}}$; and (4) BPRMF-NN with the neural network-based PLV using $P_{\mathrm{neural}}$. Table~\ref{tab:performance_ablation} shows the RMSE for the five different inputs. Overall, all models that control for PLV outperform the base model that does not control for it, although we do not see a performance gain that is statistically significant based on Welch's $t$-test at the $p < 0.05$ level. In sum, the ablation study confirms the effectiveness of IPS reweighting for modeling PLV. In the following experiments, we will use the debiasing framework with the virality-based PLV.

 % Comparing BPRMF vs. BPRMF-V, we do not see a performance gain from the BPRMF with the biased PLV that is statistically significant based on Welch's $t$-test at the $p < 0.05$ level.

\begin{table}[ht]
    \centering
    \footnotesize
    \begin{tabular}{lclc}
    \toprule
        &   User & & \\
        \multirow{-2}{*}{Model} & attributes & \multirow{-2}{*}{PLV} & \multirow{-2}{*}{RMSE} \\
        
    \midrule
        Base & \includegraphics[width=2mm, height = 2mm]{figures/tick-icon.png} & --- & 0.0866 \\
        % BPRMF & \includegraphics[width=2mm, height = 2mm]{figures/tick-icon.png} & Biased & \textbf{0.0861}\\
        BPRMF-V & \includegraphics[width=2mm, height = 2mm]{figures/tick-icon.png} & Virality-based propensity & \textbf{0.0863}\\
        BPRMF-F & \includegraphics[width=2mm, height = 2mm]{figures/tick-icon.png} & Follower-based propensity & 0.0864\\
        BPRMF-NN & \includegraphics[width=2mm, height = 2mm]{figures/tick-icon.png} & NN-based propensity & 0.0865\\
    \bottomrule
    \end{tabular}
    \caption{Ablation study using different inputs for the past latent vulnerability. Results are averaged over 5 runs.}
    \label{tab:performance_ablation}
\end{table}

\section{Discussion}
{\textbf{Summary of findings:} In our paper, we show the effects of user attributes on hate speech resharing while controlling for the past latent vulnerability of a user to hate speech. We find that users with fewer posts, fewer followers, and fewer friends tend to reshare more hate speech and newer accounts tend to reshare less. When splitting hate speech into different types, we find that racist and misogynistic rants are spread mostly by newer accounts, accounts with fewer posts, accounts with fewer friends, and accounts with a moderate number of friends. In return, political anti-Trump and anti-right-wing rants are reshared by older accounts, accounts with a moderate number of posts, and accounts with many followers and friends. Toxic rants are also shared by older accounts and accounts with a large number of posts, followers, and friends. }

{\textbf{Interpretation:} A possible explanation for our findings might be the concept of social influence. Social influence is one of the most important determinants in explaining the spread of information on social media \cite{Zaman.2014}. Users with many followers tend to have more social influence, and, therefore, their posts reach a large audience \cite{Suh.2010, Prollochs.2023, Maarouf.2024, Vosoughi.2018, Zaman.2014}. Similarly, verified users are perceived to have more credibility online \cite{Morris.2012} and, hence, have higher social influence. Users with a large social influence (e.g. verified, older accounts, more posts, more friends, and more followers) might be more concerned about having a positive image \citep{Talwar.2019}, resulting in a lower probability to reshare harmful content in general.} 

{When looking at the different types of hate speech, we make the interesting observation that the dynamics behind the of spread racial and misogynistic hate and political hate differ. Users with little social influence tend to spread more racial and misogynistic hate while users with high social influence tend to spread political hate. This might be the case because racial and misogynistic hate is socially unacceptable and circulates primarily in small online communities where extreme views are normalized and users do not have to fear backlash for resharing. In return, the political landscape is widely discussed online and political hate might border what is still deemed acceptable to express online or is harder to detect for moderation efforts.  }

% Relevance
\textbf{Relevance:} 
Regulatory initiatives in many countries around the world (e.g., the Digital Services Act in the European Union) compel social media platforms to limit the spread of online hate speech. In this regard, our framework helps platform providers identify which user attributes make users reshare hate speech. This allows for the adoption of a user-centric design approach when developing targeted and data-driven mitigation strategies. By tailoring interventions to address the roots of hate speech resharing, we can implement more effective measures to prevent users from sharing hate speech and, thereby, counter its spread.

% Implications
\textbf{Implications:} Our framework shows the influences of user attributes on the probability to reshare hate speech and therefore generates insights on how to take action against the resharing of hate speech. For example, online platforms might consider not showing the popularity of a user by removing follower counts from the profile information. This way, users are unaware of how many people see their posts, which might lead them to be more selective in the content they share. 

Platforms could also vary their moderation measures depending on the user's follower count, e.g., decrease the response times to reports against hate speech from popular users. However, this would require a shift in resources and could raise ethical considerations about the uneven moderation of users. Given the positive effects of a verified status on hate speech resharing, online platforms might also consider removing the verified status. This might lead to users being more responsible in the content they reshare.

% summary of findings
\textbf{Conclusion:}
In this work, we leverage a state-of-the-art debiasing framework to model the effects of user attributes on the probability to reshare hate speech while controlling for the past latent vulnerability to hate speech. Our framework provides flexible and explainable results while achieving state-of-the-art performance. We also cluster hate speech into its multiple types, which allows us to model the resharing dynamics individually.

% \section*{Acknowledgements}
% Funding by the German Research Foundation (Grant: 543018872) is acknowledged.

\section*{Author contributions}
All authors contributed to conceptualization, results interpretation, and manuscript writing. First author contributed to data analysis. All authors approved the manuscript.
\bibliographystyle{ACM-Reference-Format}
\bibliography{literature}

\newpage
\newpage

\appendix

\section{Statistics of user attributes}
In our work, we estimate the effects of five user attributes on the probability to reshare: verified (= 1 if the user is verified,
0 otherwise); account age (in days); \#post (number of posts
by the user); \#follower (number of followers of the user); and
\#friend (number of friends of the user). The below Table~\ref{tab:descriptives_user_attributes} shows the summary statistics of the user attributes. 

\begin{table}[ht]
    \centering
    \footnotesize
    \begin{tabular}{lrrr}
    \toprule
        User attributes & Mean & Median & Std. dev. \\
    \midrule
        verified & 0.003 & 0 & 0.05\\
        account age (in days) & 3663.74 & 3585 & 833.51 \\
        \#post & 57588.09 &  24465.5 & 106129.73\\
        \#follower & 2230.27 & 546 & 33567.04 \\
        \#friend & 1253.2 & 491 & 3639.85\\
    \bottomrule
    \end{tabular}
    \caption{Summary statistics of the user attributes.}
    \label{tab:descriptives_user_attributes}
\end{table}

\section{Hyperparameter tuning}
\label{sec:hyperparam}

Hyperparameters are tuned via cross-validation with a 70/30 split. Table~\ref{tab:hyperparam} shows the list of hyperparameters of \textbf{DF-EBM}. We fine-tuned the \textbf{DF-EBM} with the BPRMF-V confounder and applied the best-performing parameters to all our variants of \textbf{DF-EBM} model. As a result, we set the learning rate to 0.01, the number of maximum bin to 512, and the minimum sample leaf to 3.

\begin{table}[ht]
    \centering
    \footnotesize
    \begin{tabular}{lll}
    \toprule
        \textbf{Model} & \textbf{Hyperparameter} & \textbf{Search range} \\
    \midrule
         & Learning rate & 0.1, 0.01, 0.001 \\
        & Maximum bins & 128, 256, 512 \\
        \multirow{-3}{*}{DF-EBM}& Minimum sample leaf & 1, 2, 3 \\
        \bottomrule
    \end{tabular}
    \caption{Hyperparameters and search ranges for \textbf{DF-EBM}.}
    \label{tab:hyperparam}
\end{table}

\section{Insights into the different types of hate speech}
\label{sec:app_clusters}
We cluster 270 hate speech root posts into four types using unsupervised machine learning and automatically label them using a large language model. Table~\ref{tab:cluster_words} shows the accumulative number of reshares for posts in each cluster as well as the top-10 characteristics words in each cluster. We observe that clusters \say{Racist and misogynistic rants}, \say{Anti-Trump and anti-right wing rants}, and \say{Toxic rants} contain a multitude of offensive and swear words as characteristics words while the smallest cluster of social media noise does not. Moreover, Table~\ref{tab:example_tweets} shows example posts from each cluster.

\begin{table}[]
    \centering
    \footnotesize
    \begin{tabularx}{\linewidth}{>{\raggedright}p{0.25\linewidth} p{0.13\linewidth} p{0.5\linewidth}}
    \toprule
         \textbf{Hate speech cluster} & \textbf{Reshares} & \textbf{Characteristic words}  \\
    \midrule
         Racist and misogynistic rants & 15.219 & hate, niggas, nigga, dont, yall, like, ass, ugly, bitch, fucking \\
    \midrule
         Anti-Trump and anti-right wing rants & 6.207 & idiot, trump, isis, idiots, syria, says, attack, obama, war, run \\
    \midrule
         Toxic rants & 512 & dont, women, fucking, brutal, disgusting, life, white, just, bad, lifequotes \\
    \midrule
         Social media noise & 51 & summer, tomorrow, selby, model, focus, views, hes, make, time, just \\
    \bottomrule
    \end{tabularx}
    \caption{Number of reshares and characteristic words of each hate speech cluster defined as the top-10 words with regard to c-TF-IDF weighting.}
    \label{tab:cluster_words}
\end{table}

\begin{table}[]
    \centering
    \footnotesize
    \begin{tabularx}{\linewidth}{>{\raggedright}p{0.25\linewidth} p{0.68\linewidth}}
    \toprule
        \textbf{Hate speech cluster} & \textbf{Example posts} \\
    \midrule    
        Racist and misogynistic rants & \tiny$\bullet$\footnotesize\say{\emph{I understand why niggas dont wanna be in relationships girls annoying af}} \\
        & \tiny$\bullet$\footnotesize\say{\emph{I hate Niggaas that act like bitches}} \\
        & \tiny$\bullet$\footnotesize\say{\emph{RT if you still hate this nigga}} \\
    \midrule
        Anti-Trump and anti-right wing rants & \tiny$\bullet$\footnotesize\say{\emph{Join us today on the DC mall for Pussy March 2 against evil fascist dictator Drumpf. \#RESISTANCE \#NOTMYPRESIDENT \#IMWITHHER}} \\
        & \tiny$\bullet$\footnotesize\say{\emph{Breathtaking.....making money from the Presidency and his idiot supporters don't see a problem....}} \\
        & \tiny$\bullet$\footnotesize\say{\emph{RT if you're sick of these Nazis \#trumprussia}}  \\
    \midrule
        Toxic rants & \tiny$\bullet$\footnotesize\say{\emph{Dasani water is fucking disgusting and if you don't agree I literally can't even with you}} \\
        & \tiny$\bullet$\footnotesize\say{\emph{I wanna beat his ass so bad}} \\
        & \tiny$\bullet$\footnotesize\say{\emph{People are dumb af in these Jurassic Park movies! Just screaming knowing damn well not to because it attracts attention!!}} \\
    \midrule
        Social media noise & \tiny$\bullet$\footnotesize\say{\emph{They've booked shows for Efe even when he's still in the house.And some idiot who's just model in payporte said he's talentless  \#BBNaija}} \\ 
        & \tiny$\bullet$\footnotesize\say{\emph{Last day to make this a reality- Don't be fooled by the countdown, every donation still makes a HUGE difference!}}  \\
        & \tiny$\bullet$\footnotesize\say{\emph{Lovely little cove, time for some holiday painting @USER you would be proud!}} \\
    \bottomrule

    \end{tabularx}
    \caption{Example posts from each hate speech cluster}
    \label{tab:example_tweets}
\end{table}

\section{Robustness checks}
\label{sec:robustness_checks}

\textbf{Modeling past latent vulnerability:} We perform a robustness check to validate our choice of how we model the past latent vulnerability (PLV). Recall that we model PLV using the recommendation algorithm BPRMF. However, we have multiple ways to debias propensity scores, implying that we can use multiple inputs to the BPRMF model. Here, we check the prediction performance of the BPRMF with different propensity score inputs. We measure the prediction performance using recall@$k$ and NDCG@$k$. The former measures how many of the predicted posts were interesting to the user out of those that the model predicted to be interesting as the top-$k$ posts. The latter measures the accuracy of the prediction by comparing the ground-truth interestingness to the predicted ranking of posts for the top-$k$ ranked posts. This way, we can get an estimate of how well the recommendation algorithm was able to model PLV for each user. 

Table~\ref{tab:perf_recom_bprmf} shows the prediction performance of BPRMF with different inputs. We find that the BPRMF algorithms with debiased propensity scores are better at predicting hate speech that the user is vulnerable to. This confirms our choice to employ IPS reweighting in our debiasing framework when modeling PLV.

\begin{table}[ht]
\footnotesize
    \begin{subtable}[c]{0.5\textwidth}
    \centering
    \begin{tabular}{lllll}
    \toprule
    & \multicolumn{4}{c}{recall@$k$} \\
    \cmidrule(lr){2-5}

    \multirow{-2}{*}{Model}    & $k=20$ & $k=40$ & $k=60$ & $k=80$ \\
    \midrule
    BPRMF & 19.35 & 30.90 & 36.20 & 40.17 \\
    BPRMF-V & 18.02 & 31.18 & 36.76 & 40.69 \\
    BPRMF-F & \textbf{19.48} & \textbf{32.11}* & \textbf{37.92}* & \textbf{42.16}* \\
    BPRMF-NN & 19.23 & 31.16 & 36.62 & 40.80\\
    \bottomrule
    \end{tabular}
    \subcaption{Performance evaluated based on recall@$k$}
    \end{subtable}
    \begin{subtable}[c]{0.5\textwidth}
    \centering
        \begin{tabular}{lllll}
        \toprule
        & \multicolumn{4}{c}{NDCG@$k$} \\
        \cmidrule(lr){2-5}
        \multirow{-2}{*}{Model}    & $k=20$ & $k=40$ & $k=60$ & $k=80$ \\
        \midrule
        BPRMF & 7.02 & 9.39 & 10.33 & 10.97 \\
        BPRMF-V & 6.67 & 9.37 & 10.36 & 11.00 \\
        BPRMF-F & \textbf{7.04} & \textbf{9.63} & \textbf{10.66} & \textbf{11.35}*\\
        BPRMF-NN & 6.82 & 9.27 & 10.24 & 10.92 \\
        \bottomrule
        \end{tabular}
        \subcaption{Performance evaluated based on NDCG@$k$}
    \end{subtable}
\caption{Comparison of the prediction performance (measured in recall@$k$ and NDCG@$k$) of BPRMF when modeling PLV with different propensity score inputs. An asterisk * indicates that the improvement measured with Welch's $t$-test \cite{Welch.1947} is statistically significant at $p < 0.05$ level.}
\vspace{-0.3cm}
\label{tab:perf_recom_bprmf}
\end{table}

\textbf{Robustness check for smoothing operator}
In IPS for reweighting observational social media data, the IPS for rare events could be very high. A smoothing operator can lead to more stable propensity scores and prevent overfitting. In this work,  $\mu$ is the smoothing operator. We chose $\mu = 0.5$ since popularity-related measures follow a power law distribution, i.e., $\mu \leq 1$. We performed a new robustness check to study the sensitivity of the recommendation algorithm with $\mu = [0.1, 0.5, 1.0]$ for the virality and the follower-based propensity score. The results with respect to recall@$k$ are in Table~\ref{tab:robustness_mu}. We found that the overall findings remain robust.

\begin{table}[ht]
    \centering
    \footnotesize
    \begin{tabular}{lcccc}
    \toprule
        & \multicolumn{4}{c}{recall@$k$} \\
    \cmidrule(lr){2-5}

    \multirow{-2}{*}{Model}    & $k=20$ & $k=40$ & $k=60$ & $k=80$ \\
    \midrule
        BPRMF-V $\mu=0.1$ &	19.41 &	31.47 &	36.40 & 40.50 \\
        BPRMF-V $\mu=0.5$ &	18.02 &	31.18 &	36.76 &	40.69 \\
        BPRMF-V $\mu=1.0$ &	18.56 &	31.08 &	36.72 &	40.53 \\
        BPRMF-F $\mu=0.1$ &	18.73 &	30.10 &	35.28 &	39.32 \\
        BPRMF-F $\mu=0.5$ &	19.48 &	32.11 &	37.92 &	42.16 \\
        BPRMF-F $\mu=1.0$ &	18.52 &	31.74 &	37.04 &	41.14 \\
    \bottomrule

    \end{tabular}
    \caption{Performance comparison of BPRMF with varying degrees of $\mu$.}
    \label{tab:robustness_mu}
\end{table}

\section{Comparison of the past latent vulnerability to normal and hate speech}
\label{sec:comparison_normal_content}
To see whether PLV to hate speech is different from PLV to normal content, we apply the above recommendation model to users who share normal content. We then compare the PLV embeddings of hate speech and normal content by visualizing the differences with t-SNE \cite{vanderMaaten.2008}. In addition, we cluster the hate speech and normal content embeddings using DBSCAN \cite{Ester.1996} and compute the silhouette score. The silhouette score ranges from $-1$ to $+1$ and measures the separability of clusters.

Overall, we observed the following. In the t-SNE plot, we show that PLV embeddings of normal content are mostly concentrated in one broad cluster whereas the embeddings of hate speech are separated into multiple, clearly defined clusters (see Figure~\ref{fig:tsne_scatters}). We also find slightly higher silhouette scores for hate speech than normal content (0.609 and 0.539, respectively) when keeping the DBSCAN parameters constant (eps $= 0.5$ and minimum samples $= 10$). This suggests that PLV to hate speech is more homogenous within clusters and heterogenous across clusters, whereas PLV to normal content cannot be clustered into clearly distinct groups as well. This is in line with previous research which states that hate speech spreaders are more similar to each other \cite{Beatty.2020, Evkoski.2022} and interact in highly connected networks \cite{Maarouf.2024, Mathew.2018}.

\begin{table}[ht]
    \centering
    \begin{tabularx}{0.5\textwidth}{XX}
       \includegraphics[width=0.23\textwidth]{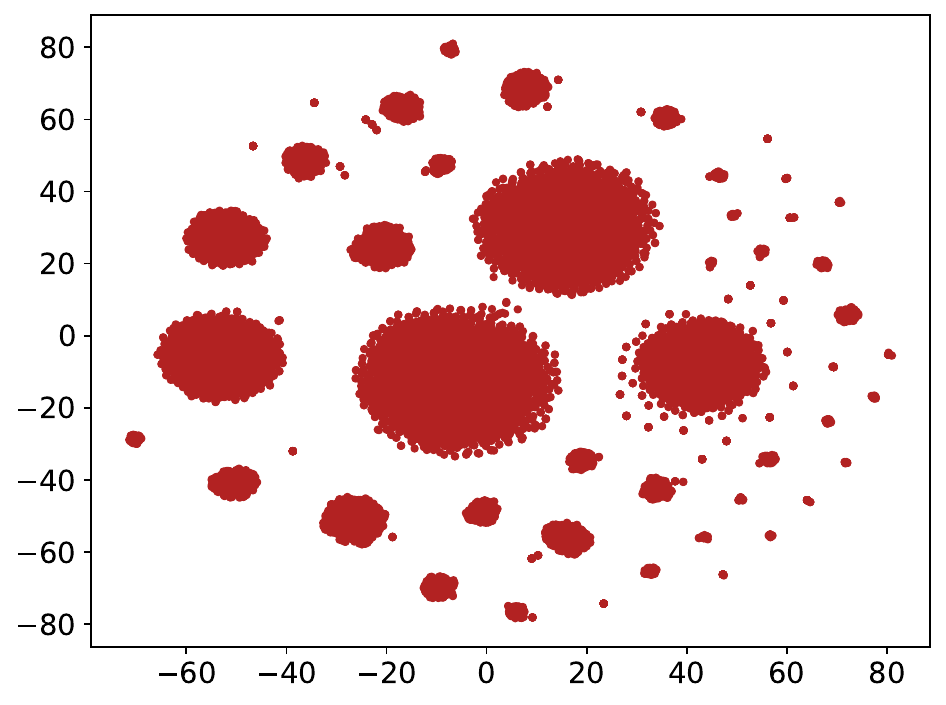}  & \includegraphics[width=0.23\textwidth]{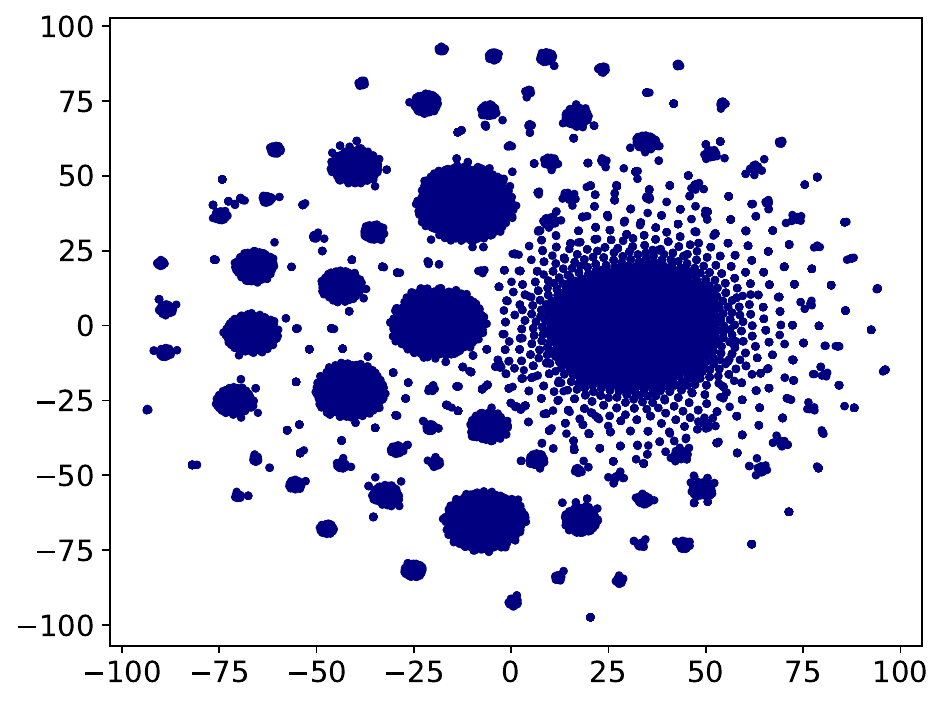} \\
       \textbf{(a) Past latent vulnerability to hate speech (Silhouette score~$=0.609$)} & \textbf{(b) Past latent vulnerability to normal content (Silhouette score~$=0.539$)}\\
    \end{tabularx}
    \caption{Comparison of the past latent vulnerability for (a)~hate speech and (b)~normal speech using a 2D $t$-SNE plot \cite{vanderMaaten.2008}.}
    \label{fig:tsne_scatters}
\end{table}

\end{document}